\documentclass[a4paper,11pt]{article}
\usepackage{jcappub} 
\usepackage{lineno}
\newcommand{\nn}{\nonumber}
\newcommand{\corr}[1]{\langle #1 \rangle}

\title{\boldmath Curvature perturbations from preheating with scale dependence}

\author{Pulkit S. Ghoderao}
\author{and Arttu Rajantie}
\affiliation{Department of Physics, Imperial College London, London SW7 2AZ, UK}

\emailAdd{pulkit.ghoderao18@imperial.ac.uk, a.rajantie@imperial.ac.uk}

\abstract{We extend the formalism to calculate non-Gaussianity of primordial curvature perturbations produced by preheating in the presence of a light scalar field. The calculation is carried out in the separate universe approximation using the non-perturbative delta N formalism and lattice field theory simulations. Initial conditions for simulations are drawn from a statistical ensemble determined by modes that left the horizon during inflation, with the time-dependence of Hubble rate during inflation taken into account. Our results show that cosmic variance, i.e., the contribution from modes with wavelength longer than the size of the observable universe today, plays a key role in determining the dominant contribution. We illustrate our formalism by applying it to an observationally-viable preheating model motivated by non-minimal coupling to gravity, and study its full parameter dependence. 
}

\begin{document}
\maketitle
\flushbottom

\section{Introduction}

Observations of the cosmic microwave background radiation and large scale structure provide strong evidence for a period of inflation in the early universe, but they do not distinguish well between different specific models of inflation. One reason for this is that in typical models the inflationary era itself is very simple, consisting of slowly rolling scalar fields, and therefore the observational predictions can also be parameterised by a small number of slow-roll parameters. Because of this it is interesting to consider the physics of reheating at the end of inflation, which can be very different in different models.

In particular, in many inflationary models reheating involves rapid particle production caused by a parametric resonance between the inflaton and other fields, commonly referred to as preheating. Preheating can be described as the stage after the end of inflation when field inhomogenities grow exponentially leading to large occupation numbers, which then back-react due to non-linear terms in the equation of motion causing the growth to stop at a point where universe enters radiation domination. Large occupation numbers mean this stage can be treated semi-classically and is equivalent to solving the full equations of motion numerically with initial conditions that are obtained from the end of inflation \cite{Polarski:1995jg, Khlebnikov:1996mc}. 

In the presence of light scalar fields, preheating produces curvature perturbations on observable scales~\cite{Chambers:2007se}. Lattice field theory simulations have shown that they can have observable levels of non-Gaussianity, especially in the massless preheating model \cite{Chambers:2007se, Chambers:2008gu, Bond:2009xx, Imrith:2019njf}. However, this model is incompatible with tensor-to-scalar ratio measurements from Planck satellite \cite{Planck:2018jri}. 
In this paper, we address this issue by considering an observationally viable model of preheating with the inflaton $\phi$ non-minimally coupled to gravity, that decays into a massless spectator field $\chi$. 

The main highlight of our article is that we perform a full calculation by including time-dependence of Hubble rate until the end of inflation in order to obtain initial conditions for numerical simulations. Such a dependence was previously considered in Ref.~\cite{Chambers:2007se} for the massless preheating model. However, as shown by Ref.~{\cite{Bond:2009xx}} the dependence of scalar curvature perturbations on the spectator field value constitutes a spiky pattern, attributable to chaotic behaviour during preheating. This non-linear behaviour requires use of a non-perturbative treatment for obtaining the non-Gaussianity from scalar curvature perturbations \cite{Suyama:2013dqa,Imrith:2018uyk}. 

We find that including the time-dependence of Hubble rate places tight constraints on the ``cosmic variance", the values which the mean spectator field can take. This is demonstrated for the preheating model we consider, but would in general be true for any other model of inflation and reheating. A negligible cosmic variance implies the mean spectator field value lies close to zero. As many inflationary potentials are symmetric around zero, it then implies the leading non-perturbative result would be the one at higher order in the non-perturbative delta N formalism. We calculate this result keeping in mind the scale dependence, showing that a simple scale-invariant way of solving momentum integrals \cite{Boubekeur:2005fj} is insufficient to yield the correct non-Gaussianity from preheating.  

The article is divided as follows: 
In Section~\ref{dN}, we begin by introducing curvature perturbations arising from preheating and the delta N formalism used to extract non-Gaussianity from them. Then we introduce and extend the non-perturbative delta N formalism to include scale dependence. These are the main results of our article. Next, in Section~\ref{sec:preheating}, we build an observationally viable model of preheating motivated by non-minimal coupling to gravity. In Section~\ref{sec:inicond}, we consider the time-dependence of Hubble rate to obtain initial conditions for simulations, and find that cosmic variance plays an important role in determining non-Gaussianity. Lastly, in Section~\ref{sec:results}, we present the $f_\text{NL}$ calculation from lattice simulations for our preheating model including its full parameter dependence.

\section{Curvature perturbations from preheating} \label{dN}

\subsection{Non-Gaussian curvature perturbations}
The primordial curvature perturbation $\zeta$ is one of the central observables in cosmology. On large scales, it can be measured directly as the relative temperature anisotropy of the cosmic microwave background radiation, and on smaller scales it acts as the seed for structure formation. Assuming that the universe is statistically homogeneous and isotropic, the curvature perturbation can be characterised by its correlation functions in momentum space. The two-point and three-point correlation functions define the curvature power spectrum $P_\zeta$ and bispectrum $B_\zeta$, respectively,
\begin{align}
\corr{\zeta(\vec{k}) \zeta(\vec{k}')} &= (2\pi)^3 \delta^{(3)}(\vec{k}+\vec{k}')P_\zeta(\vec{k}),\\
\corr{\zeta(\vec{k}_1)\zeta(\vec{k}_2)\zeta(\vec{k}_3)} &= (2\pi)^3 \delta^{(3)}(\vec{k}_1+\vec{k}_2+\vec{k}_3) B_\zeta(\vec{k}_1,\vec{k}_2,\vec{k}_3).
\end{align}
It is also convenient to define the power spectrum per logarithmic scale
\begin{equation}
\label{eq:curlyP}
	{\cal P}_\zeta(k)=\frac{k^3}{2\pi^2}P_\zeta(k).
\end{equation}
This has been measured to be
\begin{equation}
	{\cal P}_\zeta(k_*)=2.1\times 10^{-9}\equiv{\cal P}_*
\end{equation}
at the comoving scale $k_*$ corresponding to physical scale $k_\text{phys}=0.05~{\rm MPc}^{-1}$~\cite{Planck:2018nkj}. 

If the curvature perturbation is exactly Gaussian, the bispectrum $B_\zeta$ vanishes. Therefore
	the level of non-Gaussianity can be parameterised by the ratio~\cite{Komatsu:2001rj}
\begin{align}\label{eq:fNLdef}
	f_\text{NL}(\vec{k}_1,\vec{k}_2,\vec{k}_3) = - \frac{5}{6} \left(\frac{B_\zeta(\vec{k}_1,\vec{k}_2,\vec{k}_3)}{P_\zeta(\vec{k}_1) P_\zeta(\vec{k}_2) + P_\zeta(\vec{k}_1) P_\zeta(\vec{k}_3) + P_\zeta(\vec{k}_2) P_\zeta(\vec{k}_3)}\right).
\end{align}
In general, this is a function of the momenta, but in many cases it is approximately constant. This constant value is referred to as ``local'' $f_\text{NL}$, and it is constrained by the Planck observations as $f^\text{local}_\text{NL} = -0.9 \pm 5.1$~\cite{Planck:2019kim}. 

Simplest inflation models based on a single slowly-rolling inflaton field predict very highly Gaussian curvature perturbation, typically $f_\text{NL}< 0.1$, which is well below observational bounds. Therefore a detection of a definite non-zero $f_\text{NL}$ would rule out those models. Conversely, any theory that predicts significant $f_\text{NL}$ can be tested and constrained by current and future observations.

However, it is not necessarily the case that only the inflaton field generates curvature perturbations. They can equally be generated by non-adiabatic fluctuations in another scalar field, which get converted to adiabatic fluctuations during super-horizon evolution \cite{Lyth:2001nq}. Such a field is known as a curvaton. The curvaton can be thought of as taking away the burden from the inflaton to both generate inflation and curvature. This in turn allows for a wide variety of inflaton origin theories such as axion inflation \cite{Enqvist:2001zp}, string inflation \cite{Moroi:2001ct} and others to become viable.     
 
 It was noted quite early on \cite{Kofman:1994rk} that the period when energy from inflaton is transferred to Standard Model particles known as reheating \cite{Allahverdi2010} occurs in a non-linear manner. Massless preheating \cite{Kofman1997} was the toy model with potential,
 \begin{align}
 \label{equ:masslesspre}
 V(\phi,\chi) = \frac{\lambda}{4} \phi^4 + \frac{1}{2} g^2 \phi^2 \chi^2,
 \end{align}
 used to study reheating between inflaton $\phi$ and a scalar field $\chi$ \cite{Zibin:2000uw}. Here, the second field $\chi$ can also be thought of as a massless curvaton. During preheating, energy is rapidly transferred from the inflaton to the spectator field via parametric resonance. 

 In terms of perturbation theory, both fields can be split into a background part and a fluctuation. During inflation the background $\chi$ field is negligible. Inflation ends when the background $\phi$ field reaches near its minimum and slow-roll condition fails. However, due to the build up of kinetic energy, it overshoots the minimum and begins oscillating around it. The linear order equations of motion for the fluctuations of both fields now have an oscillating potential arising from the background $\phi$. It is a generic feature of second order differential equations with an oscillating term that their solutions can develop an instability for particular parameter values, and grow exponentially. For the massless preheating case, parametric resonance occurs between $g^2/\lambda = 1$ and $g^2/\lambda = 3$ for the zero momentum mode \cite{Kofman1997}. 

 In reality, fluctuations cannot grow indefinitely and stop when non-linear terms in their equations of motion start to become significant. After a sufficient time has passed, the fluctuations die down and the background inflaton settles into its minimum. With massless preheating (\ref{equ:masslesspre}), since both fields are massless, the universe then exits into a radiation dominated equation of state. The entire evolution from end of inflation to onset of radiation domination can be captured fully by using lattice simulations \cite{Chambers:2007se,Chambers:2008gu,Bond:2009xx} which predict large non-Gaussianity for the massless preheating model. However, this model is not realistic as it predicts a tensor-to-scalar ratio that is too high \cite{Imrith:2019njf} when compared to current observations \cite{Planck:2018jri}.

\subsection{Delta N formalism}

In typical inflationary models, the primordial curvature perturbation arises predominantly from the inflaton field and is nearly Gaussian, but as was shown in Refs.~\cite{Chambers:2007se,Chambers:2008gu, Bond:2009xx}, preheating can produce a potentially observable non-Gaussian contribution.

Because we are interested in the curvature perturbations on very large, superhorizon scales, we can use the {\em separate universe} approximation to describe its evolution. 
This approximation is based on the observation that if two spatial volumes are sufficiently far apart, so that light cannot travel from one to the other during the time interval we are interested in, they evolve independently of each other. In the calculation, they can therefore be thought of as two separate universes. In practice, the size of these separate spatial volumes can be taken to be of the order of the Hubble length, and therefore we will refer to them as Hubble volumes.

Furthermore, it is often the case that the universe can be assumed to be nearly homogeneous and isotropic on the scale of one Hubble length, and in that case we can approximate each separate universe with the Friedmann-Robertson-Walker (FRW) metric,
\begin{align}
	\label{eq:FRW}
ds^2 = -dt^2 + a(t)^2 \delta_{ij} dx^idx^j.
\end{align}
In this approximation, each Hubble volume therefore has its own scale factor $a(t)$ which evolves according to the Friedmann equation
\begin{equation}
	\label{eq:Friedmann}
	\left(\frac{\dot{a}}{a}\right)^2=\frac{\rho}{3M_{\rm P}^2},
\end{equation}
where $\rho$ is the local energy density in that Hubble volume.

The curvature perturbation $\zeta$ can be defined as the logarithm of the scale factor on a uniform energy density slice. For practical calculations it is convenient to define an initial flat slice with a uniform scale factor $a_{\rm ini}$. The curvature perturbation on a slice with uniform energy density $\rho_\text{ref}$ is then given by the number of e-foldings between the two slices \cite{Sugiyama:2012tj}, 
\begin{align}\label{deltaNdef}
\zeta=\delta N\equiv N-\overline{N},
~\mbox{where}~
 N \equiv \ln\left(\frac{a(\rho_\text{ref})}{a_{\textrm{ini}}}\right),
\end{align} 
and $\overline N$ is the mean value of $N$ in the observable universe.

If the equation of state is the same everywhere, every Hubble volume expands in the same way, and therefore no curvature perturbation is generated. However, if there is a scalar field $\chi$ that has a different local value $\chi(\vec{x})$ in each Hubble volume, it can affect the amount of expansion, generating a position-dependent curvature perturbation
\cite{Salopek:1990jq},
 \begin{equation}
 	\label{eq:zetadeltaN}
 	\zeta_\chi(\vec{x}) = \delta N (\chi(\vec{x})).
 \end{equation}

In this paper, we assume that there are two scalar fields: the inflaton $\phi$ which dominates the energy density during inflation and satisfies the slow roll conditions, and another field $\chi$, which is light compared to the Hubble rate during inflation so that it has significant superhorizon correlations, and which interacts sufficiently weakly with the inflaton so that its effect can be neglected in the early stages of inflation. We also assume that its self-interactions are weak so that it can be assumed to be Gaussian.

The total curvature perturbation $\zeta=\zeta_\phi+\zeta_\chi$ also includes a contribution from the inflaton field $\phi$. 
This additive form is valid if during inflation, the inflaton field dominates the expansion and is slowly rolling and that the back-reaction from $\chi$ is negligible, as we assume. Equation (\ref{eq:zetadeltaN}) shows that to obtain the statistics of the curvature perturbation generated by the field $\chi$, one needs to determine $\delta N(\chi)$ which is a function of the local value of the field $\chi$, and the statistics of the field $\chi$.

Assuming statistical homogeneity and isotropy, the statistics of $\chi$ are fully described by the two-point correlation function
\begin{equation}
\langle \delta\chi(\vec{x}) \delta\chi(\vec{x}') \rangle =\Sigma(|\vec{x}-\vec{x}'|),
\end{equation}
or its Fourier transform $\Sigma(k)$ which satisfies
\begin{equation}
\label{eq:Sigmadef}
    \langle \delta \chi(\vec{k}) \delta \chi(\vec{k}')\rangle = \Sigma(k) (2\pi)^3\delta(\vec{k}+\vec{k}'),
\end{equation}
  where $\delta\chi=\chi-\overline{\chi}$, and $\overline\chi$ is mean value of $\chi$ in the observable universe. In analogy with Eq.~(\ref{eq:curlyP}) we also define
    \begin{equation}
        \label{equ:calP}
        {\cal P}_\chi(k)=\frac{k^3}{2\pi^2}\Sigma(k).
    \end{equation}

If $\delta N$ is not a linear function of $\chi$, then it gives rise to a non-Gaussian contribution to the curvature perturbation.
In many simple scenarios, $\delta N$ can be well approximated by a quadratic Taylor expansion,
\begin{equation}
        \label{eq:Taylor}
		\delta N(\chi) = \delta N(\overline\chi) + \delta N'(\overline\chi)\delta\chi + \frac{1}{2} \delta N''(\overline{\chi}) \delta\chi^2.
\end{equation}
Assuming that the dominant contribution to the curvature perturbation is given by $\zeta_\phi$ and that it is Gaussian, the leading term in the bispectrum is~\cite{Boubekeur:2005fj}
    \begin{equation}
        \label{eq:Bpert}
    	B_\zeta(\vec{k}_1,\vec{k}_2,\vec{k}_3)
    	= \left(\delta N'\right)^2 \delta N''\left(\Sigma(k_1)\Sigma(k_2)
    	+\Sigma(k_1)\Sigma(k_3)+\Sigma(k_2)\Sigma(k_3)\right).
    \end{equation}
Considering an equilateral configuration, $k_1=k_2=k_3=k$, Eq.~(\ref{eq:fNLdef}) then gives
    \begin{equation}
    	f_\text{NL}=-\frac{5}{6}\left(\delta N'\right)^2\delta N'' \frac{\Sigma(k)^2}{P_\zeta(k)^2}
        =-\frac{5}{6}\left(\delta N'\right)^2\delta N'' \frac{{\cal P}_\chi(k)^2}{{\cal P}_\zeta(k)^2}
        .
    \end{equation}
However, if $\delta N'$ is sufficiently small, which happens if the theory has symmetry under the sign change $\chi\rightarrow -\chi$ and $\overline\chi$ is small, the dominant term is given by
\begin{align}
    \langle \zeta(\vec{x}_1) \zeta(\vec{x}_2)\zeta(\vec{x}_3) \rangle &=
    \frac{1}{8}\left(\delta N''\right)^3\langle \delta\chi(\vec{x}_1)^2 \delta\chi(\vec{x}_2)^2 \delta\chi(\vec{x}_3)^2 \rangle 
    \nonumber\\
    &=\left(\delta N''\right)^3\Sigma(|\vec{x}_1-\vec{x}_2|)\Sigma(|\vec{x}_1-\vec{x}_3|)\Sigma(|\vec{x}_2-\vec{x}_3|)+\ldots,
\end{align}
whose Fourier transform gives
\begin{equation}
    \label{eq:pertBcubic}
    B_\zeta(\vec{k}_1,\vec{k}_2,\vec{k}_3)=\left(\delta N''\right)^3 \int \frac{d^3q}{(2\pi)^3} \Sigma(q)\Sigma(|\vec{q}+\vec{k}_1|)\Sigma(|\vec{q}+\vec{k}_1+\vec{k}_2|).
\end{equation}
In this case $f_\text{NL}$ is therefore given by~\cite{Boubekeur:2005fj}
\begin{equation}
    \label{eq:pertfNLcubic}
    f_\text{NL}=-\frac{5}{6}\frac{\left(\delta N''\right)^3}{P_\zeta(k)^2}\int \frac{d^3q}{(2\pi)^3} \Sigma(q)\Sigma(|\vec{q}+\vec{k}_1|)\Sigma(|\vec{q}+\vec{k}_1+\vec{k}_2|).
\end{equation}
However, numerical simulations~\cite{Chambers:2007se, Chambers:2008gu, Bond:2009xx,Chambers:2009ki,Imrith:2019njf} have demonstrated that in preheating scenarios, $\delta N$ can be a very complicated function and therefore this simple Taylor expansion is not applicable. Hence an improvement to this formalism in the form of \textit{non-perturbative delta N formalism}\cite{Suyama:2013dqa,Imrith:2018uyk} is needed.

\subsection{Non-perturbative delta N formalism} \label{npdeltaN}
Non-perturbative delta N formalism \cite{Suyama:2013dqa,Imrith:2018uyk} performs expansion of the curvature correlator in terms of field covariance $\Sigma(x)$ instead of expanding $\delta N$ in powers of the field $\delta\chi$ as is done in perturbative delta N approach. In general, correlation functions of the curvature perturbation $\zeta_\chi$ generated by the field $\chi$ are given by the joint probability distribution $p(\chi_1,\ldots,\chi_n)$ of field values $\chi_i\equiv\chi(\vec{x}_i)$ at points $\vec{x}_i$, $i\in\{1,\ldots,n\}$,
\begin{align}
 \label{eq:generalcorr}
 \corr{\zeta_\chi(\vec{x}_1)\cdots\zeta_\chi(\vec{x}_n)} =
 \int d\chi_1\cdots d\chi_n \delta N(\chi_1)\cdots \delta N(\chi_n) p(\chi_1,\ldots,\chi_n).
\end{align} 
Because we are assuming that the field $\chi$ is a Gaussian random field, its probability distribution can be expressed in terms of its two-point correlator $\Sigma(x)$ and its Gaussian one-point probability distribution $p(\chi)$, which is determined by its mean $\overline{\chi}$ and variance $\langle\delta\chi^2\rangle$
as
\begin{equation}
\label{equ:p}
p(\chi)=\frac{1}{\sqrt{2\pi \langle\delta\chi^2\rangle}} \exp\left(-\frac{(\chi-\overline{\chi})^2}{2\langle\delta\chi^2\rangle}\right).
\end{equation}

For example \cite{Imrith:2018uyk}, the two point correlator can be expanded as
\begin{align}\label{2ptexp}
\corr{\zeta_\chi(\vec{x}_1)\zeta_\chi(\vec{x}_2)} = \tilde{N}^2_{\chi} \Sigma_{12} + \frac{1}{2} \tilde{N}^2_{\chi\chi} \Sigma_{12}^2 + \frac{1}{4} \tilde{N}^2_{\chi\chi\chi} \Sigma^3_{12} + \frac{1}{8}\tilde{N}^2_{\chi\chi\chi\chi} \Sigma^4_{12}+ \text{Order}(\Sigma_{12}^5),
\end{align}     
where $\Sigma_{ij} = \Sigma(|\vec{x}_i-\vec{x}_j|) = \corr{\delta\chi(\vec{x}_i)\delta\chi(\vec{x}_j)}$,  and the coefficients are given by
\begin{align}\label{NPcoeff}
\tilde{N}_{\chi} &=\frac{1}{\corr{\delta\chi^2}} \int d\chi\,p(\chi) \, \delta \chi \,\delta N(\chi),\nonumber\\
\tilde{N}_{\chi\chi} &=  \frac{1}{\corr{\delta\chi^2}^2} \int d\chi\,  p(\chi) \, \delta \chi^2 \,\delta N(\chi),\nonumber\\
\tilde{N}_{\chi\chi\chi} &=  \frac{1}{\corr{\delta\chi^2}^3}  \int d\chi\,  p(\chi) \, \delta \chi^3 \,\delta N(\chi) ,\nonumber\\
\tilde{N}_{\chi\chi\chi\chi} &=  \frac{1}{\corr{\delta\chi^2}^4}  \int d\chi\,  p(\chi) \, \delta \chi^4 \,\delta N(\chi), ~~~\text{etc.}
\end{align}
The non-perturbative expansion is valid if $\Sigma_{ij} << \corr{\delta\chi^2}$.

To calculate $f_\text{NL}$, we also need the three-point correlator which can be expanded in terms of the covariance as
\begin{align}
\label{3ptexp}\nn
\corr{\zeta_\chi(\vec{x}_1)&\zeta_\chi(\vec{x}_2)\zeta_\chi(\vec{x}_3)} =\\ \nn
& \tilde{N}_{\chi}\tilde{N}_{\chi\chi}\tilde{N}_{\chi}(\Sigma_{12}\Sigma_{23} + \text{perms}) + \\ \nn
&\frac{1}{2}\tilde{N}_{\chi\chi}\tilde{N}_{\chi\chi\chi}\tilde{N}_{\chi}(\Sigma^2_{12}\Sigma_{23} + \text{perms}) + \tilde{N}_{\chi\chi}\tilde{N}_{\chi\chi}\tilde{N}_{\chi\chi}(\Sigma_{12}\Sigma_{23}\Sigma_{31}) + \\ \nn
&\frac{1}{4}\tilde{N}_{\chi\chi\chi}\tilde{N}_{\chi\chi\chi\chi}\tilde{N}_{\chi} (\Sigma^3_{12}\Sigma_{23} + \text{perms}) + \frac{1}{2}\tilde{N}_{\chi\chi\chi}\tilde{N}_{\chi\chi\chi}\tilde{N}_{\chi\chi} (\Sigma^2_{12}\Sigma_{23}\Sigma_{31} + \text{perms}) +\\ 
&\frac{1}{4}\tilde{N}_{\chi\chi}\tilde{N}_{\chi\chi\chi\chi}\tilde{N}_{\chi\chi} (\Sigma^2_{12}\Sigma^2_{23} + \text{perms}) + \text{Order}(\Sigma_{ij}^5),
\end{align}
where ``perms" indicates the sum of different permutations of $1$, $2$ and $3$.
To obtain the power spectrum and bispectrum, we need to respectively Fourier transform the two-point and three-point coordinate space correlators above.

\subsection{Scale dependence} \label{BLapp}

In Ref.~\cite{Imrith:2018uyk} the non-perturbative delta N expansion Eq.~\eqref{2ptexp} and Eq.~\eqref{3ptexp} were truncated at first order in covariance to yield power spectrum
\begin{equation}
    P^\chi_{\zeta}(k) = \tilde{N}^2_{\chi}\Sigma(k),
\end{equation}
and the bispectrum
\begin{equation}
        B^{\chi}_{\zeta}(\vec{k}_1,\vec{k}_2,\vec{k}_3) = \tilde{N}_{\chi}^2\tilde{N}_{\chi\chi}\left(\Sigma(k_1)\Sigma(k_2) + \Sigma(k_1)\Sigma(k_3)+ \Sigma(k_2)\Sigma(k_3)\right),
\end{equation}
which can be seen to be analogous to Eq.~(\ref{eq:Bpert}), with the substitutions $\delta N'\rightarrow \tilde{N}_\chi$ and $\delta N''\rightarrow \tilde{N}_{\chi\chi}.$
Therefore the value of the non-Gaussianity parameter $f_\text{NL}$ also has the same form,
 \begin{equation} \label{firstNPdeltaN}
     f_\text{NL}=-\frac{5}{6}\tilde{N}_{\chi}^2\tilde{N}_{\chi\chi} \frac{\Sigma(k)^2}{P_\zeta(k)^2}.
 \end{equation}
However, as in the perturbative case, this is not necessarily the dominant term, especially if $\tilde{N}_\chi$ is small. Therefore we go to the next order of the expansion in powers of $\Sigma(k)$,
\begin{align}
P^\chi_\zeta(k_1)&=  \tilde{N}^2_\chi \Sigma(k_1)+ \tilde{N}^2_{\chi\chi} \int d\vec{q}~\Sigma(q) \Sigma(|\vec{k}_1 - \vec{q}|)\\
\nn
B^{\chi}_\zeta(\vec{k}_1,\vec{k}_2,\vec{k}_3) &=  \left(\Sigma(k_1)\Sigma(k_2) + \text{perms}\right) \left(\tilde{N}_\chi \tilde{N}_{\chi \chi} \tilde{N}_\chi \right) \\ \nn
&+ \left(\int d\vec{q}~ \Sigma(|\vec{k}_1 -\vec{q}|) \Sigma(q) \Sigma({k_2}) + \text{perms}\right) \left( \frac{1}{2} \tilde{N}_{\chi\chi\chi} \tilde{N}_\chi \tilde{N}_{\chi\chi} - 2 \corr{\delta\chi^2}^{-1} \tilde{N}_{\chi\chi}\tilde{N}^2_\chi  \right)\\ 
&+ \left(\int \Sigma({|\vec{q} - \vec{k}_1|})\Sigma(q)\Sigma({|\vec{k}_3+ \vec{q}|}) dq\right) \left( \tilde{N}^3_{\chi\chi} - 12 \corr{\delta\chi^2}^{-1} \tilde{N}^2_{\chi} \tilde{N}_{\chi\chi} \right).
\end{align}
Considering the case $\tilde{N}_\chi\rightarrow 0$, we therefore obtain for the equilateral case $|\vec{k}_1|=|\vec{k}_2|=|\vec{k}_3|=k$ the result
\begin{align}\label{NPfNL}
    f_\text{NL} = -\frac{5}{6} \frac{\tilde{N}^3_{\chi\chi} }
    {P_\zeta(k)^2}\int \frac{d^3q}{(2\pi)^3}\,\Sigma(q)\Sigma(|\vec{q}-\vec{k}_1|)\Sigma(|\vec{q}+\vec{k}_3|),
\end{align}
which is again analogous to the earlier result (\ref{eq:pertfNLcubic}) but is valid even when the Taylor expansion (\ref{eq:Taylor}) is not.

In Ref.~\cite{Boubekeur:2005fj}, the authors assumed a scale-invariant power spectrum for the field $\chi$,
\begin{align}
\label{equ:scaleinv}
    \Sigma(k) = \frac{2\pi^2}{k^3} \frac{H_*^2}{4\pi^2},
\end{align}
where $H_*$ is the Hubble rate at the time when the current observational scale left the horizon. With that assumption, the non-Gaussianity parameter $f_\text{NL}$ given by Eq.~(\ref{NPfNL}) at the observational scale $k_*$ becomes approximately
\begin{align}\label{BLfNL}
    f_\text{NL} \approx -\frac{5}{6} \frac{\tilde{N}^3_{\chi\chi} \mathcal{P}^3_\chi }{\mathcal{P}_*^2}\int_{L^{-1}}^{k_*} \frac{dq}{q}
    =-\frac{5}{6} \frac{\tilde{N}^3_{\chi\chi} \mathcal{P}^3_\chi }{\mathcal{P}_*^2} \ln \left(k_*L\right),
\end{align}
with an infrared cut-off $L^{-1}$ corresponding to the size of the observable universe. Therefore its $k$-dependence appears logarithmic.

However, in practice the field $\chi$ will not be exactly scale invariant because of its mass and the time-dependence of the Hubble rate during inflation. To capture the effect of the scale dependence, we approximate the field variance with a power law,
\begin{align}\label{analVar}
    \Sigma(k) \approx 2\pi^2 \frac{\mathcal{A}_\chi}{k^{3-n_\chi}},
\end{align}  
where the spectral index  $n_\chi (> 0)$ and the amplitude $\mathcal{A}_\chi$, which has energy dimensions $2 - n_\chi$, depend on the parameters of the specific inflation model
as we will discuss in more detail in Section~\ref{fNLest}. This form implies the real space correlator behaves on asymptotically long distances as $\Sigma_{ij}\sim |\vec{x}_i - \vec{x}_j|^{-n_\chi}$ and is therefore small on observable scales, justifying the non-perturbative delta N expansion. Then the integral in Eq.~(\ref{NPfNL}) gives for the equilateral case,
\begin{align} \label{scaledfNL}
    f_\text{NL} \approx 
    -\frac{5}{6} \frac{\tilde{N}^3_{\chi\chi} \mathcal{A}^3_\chi k_{*}^{2n_\chi} }{\mathcal{P}_*^2}
    \int_0^{k_{*}} \frac{dq}{q^{1-n_\chi}}
    =
    -\frac{5}{6} \frac{\tilde{N}^3_{\chi\chi} \mathcal{A}^3_\chi k_{*}^{3n_\chi} }{n_\chi \mathcal{P}_*^2}.
\end{align}
This shows that $f_\text{NL}$ has power-law dependence on the observational scale $k_*$. Because cosmological observations are carried out on comoving scales that are many orders of magnitude larger than the Hubble scale during inflation, this leads to a significant suppression of $f_\text{NL}$ if $n_\chi>0$.

\section{Preheating model}
\label{sec:preheating}

\subsection{Potential}

 Calculations of non-Gaussianity in the massless preheating model have been made by simulating full non-linear behaviour in the perturbative delta N formalism \cite{Chambers:2007se}. Similar calculations using the non-perturbative delta N formalism \cite{Imrith:2019njf} yield significant non-Gaussianity $f_\text{NL}$ but the tensor to scalar ratio is very high for inflaton power spectrum within observational bounds. 
 
 It has been shown that introduction of non-minimal coupling of inflaton to gravity is able to satisfy observational bounds on spectral index and tensor to scalar ratio \cite{Bezrukov2008, Planck:2018jri}. In order to maintain the theoretical relaxations made possible by the curvaton scenario described above, along with satisfying non-Gaussianity, spectral index and tensor to scalar ratio bounds, we need to study massless preheating with non-minimal coupling to gravity. 
 
Even if no non-minimal coupling is included in the classical action, one loop calculations made using QFT in curved spacetime automatically generate the non-minimal coupling \cite{Tagirov:1972vv}. Therefore, it is necessary to include the non-minimal coupling. Indeed, a minimal coupling would require there to be some extra symmetry which forbids the non-minimal coupling term from appearing in the action. The coupling takes a scale invariant form: $\xi \phi^2 R/2$ where $\phi$ is the inflaton field, $R$ is the scalar curvature and $\xi$ is the non-minimal coupling parameter which is dimensionless. The QFT calculations further show that the non-minimal coupling parameter runs logarithmically with energy scale but does not possess any fixed point \cite{Markkanen:2013nwa}. Thus, there is no a priori constraint on the value of the non-minimal coupling parameter. In this work, we will assume that it has a small positive value $0<\xi\ll 1/6$.

The Jordan frame action for massless preheating with only the inflaton field coupled non-minimally to gravity is taken to be
\begin{align}
S = \int d^4x~\sqrt{-g}~\left(f(\phi)R- \frac{1}{2} g^{\mu\nu} \partial_\mu \phi \partial_\nu \phi - \frac{1}{2} g^{\mu\nu} \partial_\mu \chi \partial_\nu \chi- V(\phi,\chi) \right),\\
\text{where}~~~ f(\phi) = \frac{M^2_P}{2} + \frac{1}{2} \xi \phi^2~~~ \text{and}~~~ V(\phi,\chi) = \frac{\lambda}{4}\phi^4 + \frac{1}{2}g^2\phi^2\chi^2.
\end{align}
This Jordan frame action can be converted to the Einstein frame by conformally rescaling the metric to remove the non-minimal coupling and then redefining the fields to bring their kinetic terms back to the canonical form. 

In our case, using a conformal rescaling $\tilde{g}_{\mu\nu} = (1 + \xi \frac{\phi^2}{M^2_P}) g_{\mu\nu}$ yields \cite{Kaiser:2010ps}
\begin{align}
S = \int d^4x~\sqrt{-\tilde{g}}~ \left( \frac{M^{2}_P}{2} \tilde{R} - \frac{1}{2}\tilde{g}^{\mu\nu} \mathcal{G}_{ij}\partial_\mu \phi^i \partial_\nu \phi^j - \tilde{V}(\phi,\chi) \right),
\end{align} 
where 
\begin{equation}
\tilde{V}(\phi,\chi) = \frac{M^4_P}{4f^2(\phi)} V(\phi,\chi),
\end{equation}
and $\mathcal{G}_{ij}$ is the metric in field space with components
\begin{equation}
\mathcal{G}_{\phi\phi} = \frac{1}{1+\xi \frac{\phi^2}{M^2_P}} + \frac{6 \xi^2 \frac{\phi^2}{M^2_P}}{(1+\xi \frac{\phi^2}{M^2_P})^2} ~~~\text{and}~~~\mathcal{G}_{\chi\chi} = \frac{1}{1+\xi \frac{\phi^2}{M^2_P}}~.
\end{equation}
We can check that the Ricci scalar of this field space metric is not zero at order $\xi$. Hence even with only the inflaton non-minimally coupled to gravity it is not possible to bring back canonical kinetic terms for both fields as the field space metric is not conformally flat. Furthermore it has been shown \cite{Kaiser:2010ps} that bringing back canonical kinetic terms is generically not possible if more than one field is non-minimally coupled. 

However if only a single field $\phi$ was present in the Jordan frame action, then the field space metric has only one component $\mathcal{G}_{\phi\phi}$ and a redefined field variable $\tilde\phi$ which satisfies the differential equation
\begin{equation}
    \frac{d{\tilde\phi}} {d{\phi}} = \frac{\sqrt{\xi(1+6\xi) {\phi}^2/M^2_P + 1}}{ \xi {\phi}^2/M^2_P + 1}
    \approx \frac{1} {\sqrt{\xi {\phi}^2/M^2_P + 1}},
\end{equation}
where the final form is valid for $\xi\ll 1/6$, brings back the canonical kinetic term. This gives us the relation
\begin{align}
\label{eq:phitilde}
\phi = \frac{M_P}{\sqrt{\xi}}\sinh\left(\frac{\sqrt{\xi}}{M_P} \tilde{\phi}\right),
\end{align}
and the Einstein frame potential
\begin{align}
    \tilde{V}(\tilde{\phi}) = \frac{\lambda}{4} \left(\frac{M_P}{\sqrt{\xi}} \tanh(\frac{\sqrt{\xi} \tilde{\phi}}{M_P})\right)^4~.
\end{align}
Such a potential is also motivated by $\alpha-$attractor T models of inflation \cite{Kallosh:2013hoa}.

Therefore, in order to simplify numerical simulations by not having to include non-canonical kinetic terms, we choose a potential for the two-field case as
\begin{align}\label{NMCpreheatingeq}
\tilde{V}(\tilde{\phi},\tilde{\chi}) =  \frac{\lambda}{4} \left(\frac{M_P}{\sqrt{\xi}} \tanh(\frac{\sqrt{\xi}}{M_P}\tilde{\phi})\right)^4 + \frac{g^2}{2} \tilde{\chi}^2 \left(\frac{M_P}{\sqrt{\xi}} \tanh(\frac{\sqrt{\xi}}{M_P}\tilde{\phi})\right)^2 .
\end{align}
A similar approach to preheating simulations in $\alpha$-attractor motivated potentials has been used in Ref.~\cite{Antusch:2021aiw}. Here, we have kept the identification $\phi \to (M_P/\sqrt{\xi}) \tanh(\sqrt{\xi} \phi/M_P)$ in the interaction between the two fields to preserve the flatness of the potential at large inflaton values, which maintains the desirable features of $\alpha$-attractor models. Henceforth we shall drop the tilde on the fields for simplicity, bearing in mind that they belong to the Einstein frame.

\subsection{Parameter values}

 Our non-minimally coupled (NMC) preheating model has three parameters: $\lambda, g^2$, and $\xi$. Of these, we shall see that $\lambda$ is constrained by the power spectrum observation and $\xi$ is constrained by tensor-to-scalar ratio observation.
 
 During inflation we assume the field $\chi$ is negligible, so we have effectively a single-field inflationary model with the inflaton potential
 \begin{align}
     V(\phi) = \frac{\lambda}{4} \left(\frac{M_P}{\sqrt{\xi}} \tanh(\frac{\sqrt{\xi} \phi}{M_P})\right)^4.
 \end{align} 
The first slow-roll parameter is
\begin{equation}
    \label{epsiloneq}
 \epsilon = \frac{M_P^2}{2}\left(\frac{V'}{V}\right)^2
= 
 \frac{8\xi}{\cosh^2\left(\sqrt{\xi}\frac{\phi}{M_P}\right)\sinh^2\left(\sqrt{\xi}\frac{\phi}{M_P}\right)}.
\end{equation}
Setting $\epsilon=1$, we find that inflation ends when $\phi$ reaches the value $\phi_{\rm end}$ given by
 \begin{align}
\cosh\left(\frac{2\sqrt{\xi}}{M_P} \phi_\text{end}\right) = \sqrt{1+32\xi}.
 \end{align} 
We normalise the scale factor $a$ to be unity at this time, $a_\text{end}=1$.

Using the slow-roll field and Friedmann equations, we can then relate the number of e-foldings until the end of inflation ${\cal N}=-\ln a$ to the inflaton field value,
\begin{equation}
{\cal N} = \frac{1}{16\xi} \cosh\left(2\frac{\sqrt{\xi}}{M_P}\phi \right) - \frac{\sqrt{1+32\xi}}{16\xi},
\end{equation}
and the Hubble rate to ${\cal N}$ as
 \begin{align}\label{Heq}
 H &= \sqrt{\frac{\lambda}{12}} \frac{M_P}{\xi} \tanh^2\left(\frac{\sqrt{\xi}}{M_P} \phi\right) =\sqrt{\frac{\lambda}{12}} \frac{M_P}{\xi} 
 \frac{16\xi {\cal N} +\sqrt{1+32\xi}- 1}{16 \xi {\cal N} + \sqrt{1+32\xi} +1}.
 \end{align}
 The curvature power spectrum due to the inflaton field is therefore~\cite{lyth_liddle_2009}
 \begin{equation}\label{Pstareq}
 \mathcal{P}_* = \frac{1}{8\pi^2} \frac{H(\phi_*)^2}{M^2_P \epsilon(\phi_*)}= \frac{1}{768\pi^2} \frac{\lambda}{\xi^3} \frac{\sinh^6\left(\sqrt{\xi}\frac{\phi_*}{M_P}\right)}{\cosh^2\left(\sqrt{\xi}\frac{\phi_*}{M_P}\right)},
 \end{equation}
 where we have used Eqs.~\eqref{epsiloneq} and \eqref{Heq},
 and $\phi_*$ is the field value $\mathcal{N}_*$ e-foldings before the end of inflation, when the observational scale $k_*$ exited the horizon,
  \begin{align}
 \phi_* = \frac{M_P}{2\sqrt{\xi}} \cosh^{-1}(16\mathcal{N}_*\xi + \sqrt{1+32\xi}).
 \end{align}
 The precise value of $\mathcal{N}_*$ depends on post-inflationary evolution of the universe, but to be specific we will assume $\mathcal{N}_*=55$.
 Setting ${\cal P}_*$ equal to its observed value, ${\cal P}_*=2.1\times 10^{-9}$, we obtain
 \begin{align}
 \lambda = 6.4\times 10^{-5}~  \xi^3 \frac{\left(16 \mathcal{N}_* \xi+\sqrt{32 \xi+1}+1\right)}{ \left(16 \mathcal{N}_* \xi+\sqrt{32 \xi+1}-1\right)^3}.
 \end{align}
 As we can see, $\lambda$ is fully determined once we fix $\xi$.

 Parameter $\xi$ is constrained from below by the tensor to scalar ratio observation \cite{Planck:2018jri}. 
 At the observational scale $N_*$, our model gives \cite{lyth_liddle_2009}
 \begin{equation}
 r = 16 \epsilon_* = \frac{512\xi}{\sinh^2\left(2\sqrt{\xi}\frac{\phi_*}{M_P}\right)} = \frac{16}{8\mathcal{N}_*^2\xi + \mathcal{N}_*\sqrt{32\xi+1}+1}.
 \end{equation}
Requiring that $r<0.1$ for $\mathcal{N}_*=55$ gives us the lower bound
 \begin{equation}
 \label{xireq}
  \xi > 0.004.
 \end{equation}

\section{Initial Conditions}
\label{sec:inicond}
After introducing our model in the previous section, the next step is to calculate non-Gaussianity parameter $f_\text{NL}$ given by Eq.~\eqref{NPfNL}.

In order to apply the non-perturbative delta N formalism, we need to know the statistics of the $\chi$ field at the initial time slice $a_{\rm ini}$ defined in Eq.~(\ref{deltaNdef}). Because the field is weakly coupled, its evolution during inflation can be described using linear theory in momentum space.
It is convenient to divide the comoving momenta $k$ in three separate ranges: 

Modes with $k<a_0H_0$, where $a_0$ and $H_0$ are the scale factor and Hubble rate today, have wavelength longer than the size of the currently observable universe. Therefore from the perspective of any observation, they appear as a uniform, non-zero mean value, which we denote as $\overline{\chi}$. Because of this random origin, the precise value of $\overline{\chi}$ cannot be computed, but if we know the whole inflationary history, we can compute its variance $\langle\overline{\chi}^2\rangle$, which is often referred to as the ``cosmic variance".

Modes with $a_0H_0<k<a_{\rm ini}H_{\rm ini}$, where $H_{\rm ini}$ is the Hubble rate at the initial slice, have a wavelength that is shorter than the size of the observable universe today but longer than the size of a single ``separate universe". Therefore these modes give rise to a different initial value $\chi_{\rm ini}$ in each separate universe and are responsible for the produced curvature perturbation.
More precisely, in the context of the non-perturbative delta N calculation, they determine the field correlation function $\Sigma(x)$ and the one-point probability distribution (\ref{equ:p}), which depends on the mean $\overline\chi$ and the variance $\langle \delta\chi^2\rangle.$

Modes with $k>a_{\rm ini}H_{\rm ini}$, where $H_{\rm ini}$ is the Hubble rate at the initial slice, have a wavelength shorter than the size of a single ``separate universe". Therefore they are statistically homogeneous on the observational scales and do not contribute directly to the curvature perturbation.
However, they play an important role in the calculation and are discussed in Section~\ref{sec:lattice}.

At linear order, a mode $\chi_k$ with comoving momentum $k$ satisfies the mode equation
\begin{align} \label{modeeq}
\ddot{\chi}_k + 3H \dot{\chi}_k + \frac{k^2}{a^2}\chi_k + m_\chi^2(\phi) \chi_k = 0,
\end{align}
where the effective $\phi$-dependent mass of the $\chi$ field is
\begin{equation}
\label{eq:mchidef}
m_\chi^2(\phi)\equiv g^2 \left(\frac{M_P}{\sqrt{\xi}} \tanh\left(\frac{\sqrt{\xi}}{M_P}\phi\right)\right)^2.
\end{equation}
Because 
\begin{equation}
\frac{m_\chi^2}{H^2}=12\xi \frac{g^2}{\lambda} \tanh\left(\frac{\sqrt{\xi}}{M_P}\phi\right)^{-2}=
12\frac{g^2\xi}{\lambda} 
\left(1+\frac{2}{16 \xi {\cal N} + \sqrt{1+32\xi}-1} \right),
\end{equation}
we can see that assuming $g^2\xi/\lambda < 1$, the $\chi$ field is light early on during inflation, and its mass increases gradually relative to the Hubble rate during inflation. 

For convenience, we choose the initial time slice $a_{\rm ini}$ in Eq.~(\ref{deltaNdef}) to correspond to the moment when 
\begin{equation}
\frac{m_\chi^2(\phi_{\rm ini})}{H(\phi_{\rm ini})^2}=\frac{9}{4},
\end{equation}
which gives
\begin{align}\label{Nini}
\mathcal{N}_\text{ini} = \frac{1}{16\xi} \left(\frac{3+16g^2\xi/\lambda}{3 - 16 g^2 \xi/\lambda} - \sqrt{1+32\xi}\right).
 \end{align}

Because the field is light,
we can approximate that when the comoving mode $\chi_k$ exists the horizon at time $t_k$, its amplitude is
\begin{equation}\label{scaleinvPower}
    {\cal P}_\chi(k;t_k)=\frac{H(t_k)^2}{4\pi^2}.
\end{equation}
Afterwards, its evolution is overdamped and its amplitude decays with rate
\begin{equation}
    \gamma(t)=\frac{3H(t)}{2}\left[1-\sqrt{1-\frac{4m_\chi^2(t)}{9H(t)^2}}\right].
\end{equation}
At the time slice $t_{\rm ini}$, the amplitude is therefore
 \begin{align}\label{NMCPower}
 \mathcal{P}_\chi(k;t_\text{ini}) = \frac{H({\cal N}_k)^2}{4\pi^2} e^{-3F({\cal N}_k, {\cal N}_\text{ini})},
 \end{align}
 where ${\cal N}_k$ and ${\cal N}_{\rm ini}$ denote the number of e-foldings until the end of inflation at times $t_k$ and $t_{\rm ini}$, respectively, and
  \begin{align}
 F({\cal N}_k,{\cal N}) = \frac{2}{3}\int^{{\cal N}_k}_{{\cal N}} \frac{\gamma}{H}d{\cal N'}=
 \int^{{\cal N}_k}_{{\cal N}} \left(\sqrt{1 - \frac{16g^2}{3\lambda} \xi \left( \frac{16\xi {\cal N'} + \sqrt{1+32\xi} + 1}{16\xi {\cal N'} + \sqrt{1+32\xi} - 1} \right)} -1 \right)\,d{\cal N'},
 \end{align}
 which for ${\cal N}_\text{ini}$ in Eq.~\eqref{Nini} is given by
 \begin{align}
  F({\cal N}_k,{\cal N}_\text{ini}) &=     {\cal N}_k - \frac{1}{16\xi} \left(\frac{3+16g^2\xi/\lambda}{3 - 16 g^2 \xi/\lambda} - \sqrt{1+32\xi}\right) \nonumber\\
 & - \frac{16\xi {\cal N}_k + \sqrt{1+32\xi} - 1}{48\xi}\sqrt{9 - 48\frac{g^2}{\lambda}\xi \left( \frac{16\xi {\cal N}_k + \sqrt{1+32\xi} + 1}{16\xi {\cal N}_k + \sqrt{1+32\xi} - 1} \right)}
 \nonumber\\
 &+\frac{2g^2/\lambda}{\sqrt{9 - 48g^2\xi/\lambda}} \tanh^{-1}\left(\sqrt{\frac{3\lambda - 16g^2\xi\left( \frac{16\xi {\cal N}_k + \sqrt{1+32\xi} + 1}{16\xi {\cal N}_k + \sqrt{1+32\xi} - 1} \right)}{3\lambda - 16g^2\xi}} \right).
   \label{Feq}
 \end{align}
This implies the power spectrum \eqref{NMCPower} is scale dependent through the function $F(\mathcal{N}_k)$. Therefore we need to go beyond the scale-invariant approximation used in other works Refs.~\cite{Boubekeur:2005fj, Imrith:2019njf}.

In order to compute $f_{\rm NL}$ using Eq.~(\ref{NPfNL}), we need the two-point correlator $\Sigma(k)$, which is given by Eq.~(\ref{equ:calP}), as well as the one-point probability distribution $p(\chi)$, which is given by Eq.~(\ref{equ:p}) and depends on the variance $ \corr{\delta\chi^2} $ and the mean value $\overline{\chi}$.
As discussed earlier, the latter is a random variable whose variance $\corr{\overline\chi^2}$ we can compute.

The two variances are both given by a similar integral, with a different integration range,
  \begin{align} 
  \label{CVeq}
 \corr{\overline\chi^2} &= \int_0^{a_0H_0}\mathcal{P}_\chi(k;t_{\rm ini}) \frac{dk}{k}= \int^{\infty}_{\mathcal{N}_*} \frac{H({\cal N}_k)^2}{4\pi^2} e^{-3F({\cal N}_k, {\cal N}_\text{ini})}\left(\frac{1}{H({\cal N}_k)}\frac{dH({\cal N}_k)}{d{\cal N}_k}-1\right)d{\cal N}_k
 ,\\
 \label{IVeq}
 \corr{\delta\chi^2} &= \int_{a_0H_0}^{a_{\rm ini}H_{\rm ini}} \mathcal{P}_\chi(k;t_{\rm ini}) \frac{dk}{k}= \int_{\mathcal{N}_\text{ini}}^{ \mathcal{N}_*} \frac{H({\cal N}_k)^2}{4\pi^2} e^{-3F({\cal N}_k, {\cal N}_\text{ini})} \left(\frac{1}{H({\cal N}_k)}\frac{dH({\cal N}_k)}{d{\cal N}_k}-1\right)d{\cal N}_k
 ,
  \end{align}
where the latter forms are obtained by changing the integration variable to ${\cal N}_k$, the value of ${\cal N}$ at the time when mode $k$ left the horizon,
\begin{equation}
    \frac{dk}{k}=\left(\frac{1}{H({\cal N}_k)}\frac{dH({\cal N}_k)}{d{\cal N}_k}-1\right)d{\cal N}_k,
\end{equation}
with $H({\cal N}_k)$ given by Eq.~(\ref{Heq}).

Throughout this article we consistently choose the number of efolds at which the currently observable scale left the horizon to be $\mathcal{N}_* = 55$.

\begin{figure}[t]
    \centering
    \includegraphics[scale=0.25]{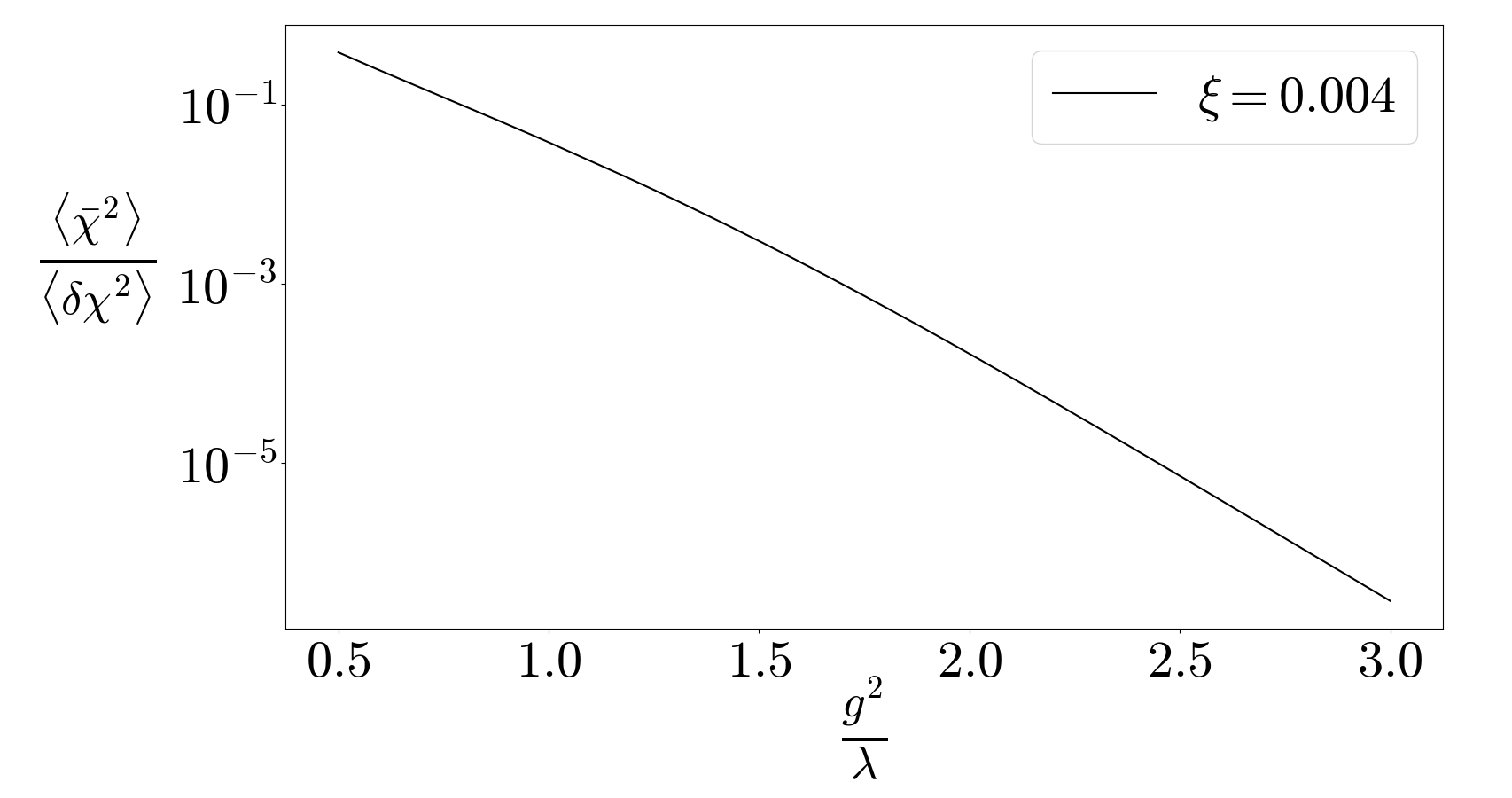}
    \caption{Ratio of the cosmic variance to the initial variance for $g^2/\lambda \in [0.5,3]$ at fixed $\xi=0.004$. Cosmic variance is two orders of magnitude less than the variance of the initial $\chi$ field in the range $1< g^2/\lambda < 3$ which is relevant for parametric resonance \cite{Kofman1997}. For $g^2/\lambda > 3$, it is more than four orders less. It only becomes comparable to the initial variance at $g^2/\lambda < 0.5$ which is well outside the parametric resonance band. 
    }
    \label{Varratio}
\end{figure}

The cosmic variance integral Eq.~\eqref{CVeq} at large $\mathcal{N}_k$ goes as
\begin{align}
    \corr{\overline{\chi}^2} \sim \int^{\infty}\exp\left(-\left(1 - \sqrt{1 - \frac{16}{3} \xi \frac{g^2}{\lambda}}\right) \mathcal{N}_k\right) d\mathcal{N}_k~. 
\end{align}
From which we may deduce that for $\xi=0$ the integral has no exponential suppression and hence cosmic variance for the minimal coupling case is infinite \cite{Chambers:2008gu}. However, the introduction of even a small positive non-minimal coupling $\xi > 0$ gives rise to an exponential suppression and hence we expect cosmic variance to be negligible for the non-minimal coupling case. This is indeed what we find by numerically integrating for the cosmic and initial variances in Eqs.~\eqref{CVeq} and \eqref{IVeq}, as summarised in Figure~\ref{Varratio}.

Whereas an infinite cosmic variance would imply that the mean $\overline{\chi}$ that appears in the probability density $p(\chi)$ can have any value and can therefore be considered to be a free parameter, a negligible cosmic variance forces us to choose $\overline{\chi} \approx 0$. Symmetry of the potential around zero dictates $\tilde{N}_\chi$ will be small which, as we saw in Section~\ref{npdeltaN}, makes the dominant term in the non-perturbative delta N formalism to be Eq.~\eqref{NPfNL}. Thus we illustrate the important role played by cosmic variance in calculating the non-Gaussianity from preheating. If cosmic variance is not negligible, as happens for the massless preheating toy model, then one needs to use Eq.~\eqref{firstNPdeltaN} as done in Ref.~\cite{Imrith:2019njf}. However if it is negligible, as happens in our NMC preheating model, then we need to use Eq.~\eqref{NPfNL} to calculate $f_\text{NL}$. The only qualification is for the potential to be symmetric in $\chi$, which is true in most preheating scenarios.

\section{Numerical results} 
\label{sec:results}
\subsection{Lattice simulations}
\label{sec:lattice}

We use the program HLattice \cite{Huang:2011gf} to perform fully non-linear simulations of preheating dynamics. For a specific value of $g^2/\lambda$ and $\xi$, we draw $N_{\rm runs}=10,000$ initial field values $\chi_i$, $i\in\{1,\ldots,N_{\rm runs}\}$, from the probability distribution (\ref{equ:p}) with zero mean, i.e., $\overline{\chi}=0$, and with $\corr{\delta \chi^2}$ given by Eq.~(\ref{IVeq}).
A lattice simulation is run and the value of $\delta N$ is extracted for each of the $N_{\rm runs}$ values separately.
This amounts to importance sampling the probability distribution, and therefore the coefficients $\tilde{N}_\chi$ and $\tilde{N}_{\chi\chi}$ defined in Eqs.~(\ref{NPcoeff}) can be obtained as averages over the runs, 
\begin{align}
\tilde{N}_\chi &= \frac{1}{N_{\rm runs}}\sum_i \left(\chi_i-\overline{\chi}\right)\left(N(\chi_i)-\overline{N}\right),
\nonumber\\
\tilde{N}_{\chi\chi} &= \frac{1}{N_{\rm runs}}\sum_i \left(\chi_i-\overline{\chi}\right)^2\left(N(\chi_i)-\overline{N}\right).
\end{align}

The fields are evolved in a linear fashion till the condition that the potential energy is equal to the kinetic energy
is met. Thereafter non-linear simulations begin. Using the principles first outlined in \cite{Khlebnikov:1996mc}, we can study the semi-classical dynamics occurring during preheating by classically simulating an equivalent non-equilibrium, non-linear problem. In HLattice, this is achieved through capturing the quantum nature by initialising the fields with random Gaussian fluctuations for inhomogeneous modes. 

Full equations of motion for the fields and Friedmann equation are solved on a comoving lattice with $32^3$ grid points and length of a side equal to $20/H$, where $H$ is the Hubble rate at the start of the non-linear evolution. The lattice size has been chosen as a compromise between two factors: Ideally, we would want the lattice to be large enough so that light cannot travel across the lattice during the nonlinear non-equilibrium stage of the evolution, but if it is larger than the comoving Hubble length, then we would need to include metric perturbations and the use of the FRW metric would not be justified. For our lattice size, the comoving Hubble length is initially somewhat smaller than the lattice size but always lies above the lattice spacing. We choose the discretisation scheme LATTICEEASY for evolution in HLattice as we are not including any metric perturbations. 

 We have also checked that changing the lattice spacing from $32^3$ to $16^3$ grid points while keeping lattice size $20/H$ fixed as well as changing the lattice spacing from $32^3$ to $64^3$ grid points while changing lattice size from $20/H$ to $40/H$ does not alter our simulation results significantly. Furthermore, the comoving mass scale $(am_\chi)^{-1}$ where $m^2_\chi = g^2 \corr{\tanh^2{\sqrt{\xi}\phi}}/\xi$ that is a representative of the field dynamics, lies between the lattice spacing and lattice size throughout the non-linear evolution.

\begin{figure}[t]
 	\centering
 	\includegraphics[scale=0.33]{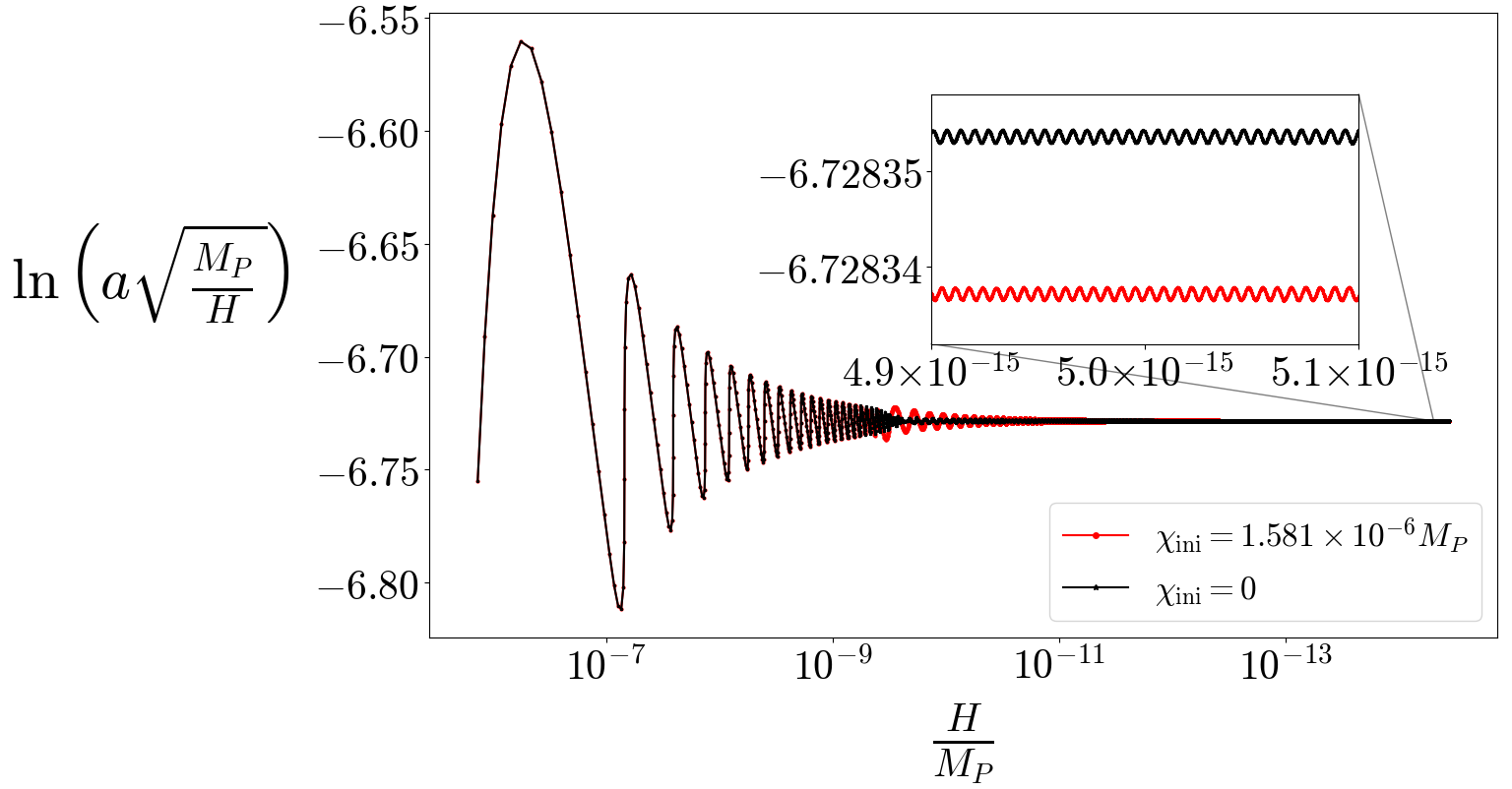}
 	\caption{The deviation from radiation domination $a \propto 1/\sqrt{H}$ at two $\chi_\text{ini}$ values for $g^2/\lambda=2$ and $\xi = 0.004$. $\chi_\text{ini} = 1.581 \times 10^{-6} M_P$ corresponds to spike in the delta N function. Inset shows zoomed in data around $H_\text{ref} = 5 \times 10^{-15} M_P$ where we pick the value of $N=\ln(a)$ after filtering.}
 	\label{peakcomp}
 \end{figure}

Plotting the raw $a$ versus $H$ data obtained from HLattice in the form of deviation from radiation domination, Figure~\ref{peakcomp} shows that all curves become flat at late times, indicating onset of radiation domination. It also shows the change in asymptote for two different values of $\chi_\text{ini}$. Using Gaussian filtering in conformal time with standard deviation of 100 time steps, we can precisely pick the value of $N=\ln(a)$ at $H_\text{ref} = 5 \times 10^{-15} M_P$ which would correspond to a constant energy density $\rho_\text{ref}$ to yield the $N$ function Eq.~\eqref{deltaNdef}. The two curves deviate due to field excursion in transverse direction \cite{Bond:2009xx}, giving rise to a spike in the delta N function at $\chi_\text{ini} = 1.581 \times 10^{-6} M_P$ as compared to $\chi_\text{ini} = 0$. 

\begin{figure}[t]
	\centering
	\includegraphics[scale=0.33]{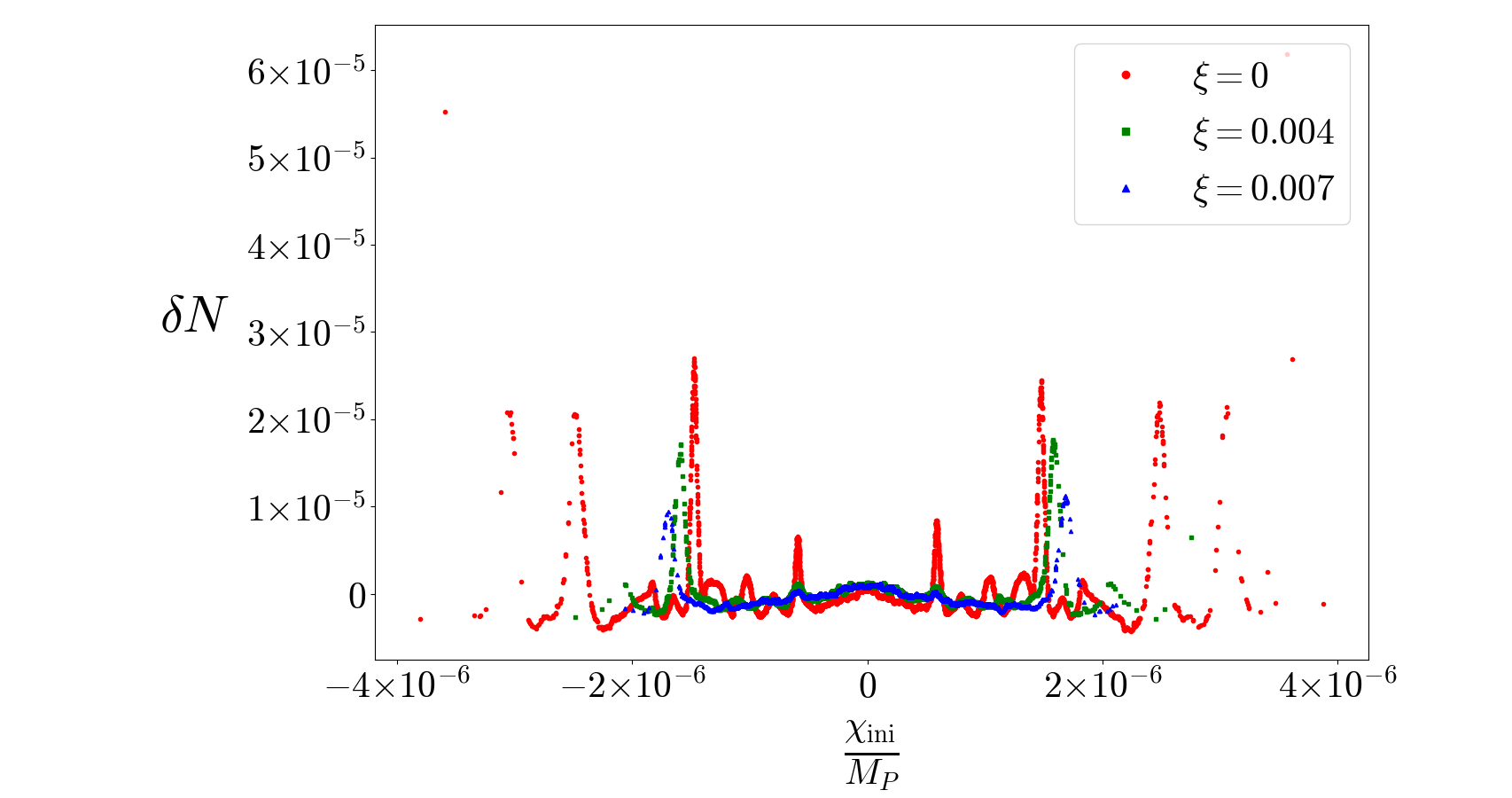}
	\caption{$\delta N$ at $H_\text{ref} = 5 \times 10^{-15}M_P$ for $\xi = 0, 0.004, 0.007$ and $g^2/\lambda=2$. The relevant initial quantities used to plot $\delta N$ are summarised in Table~\ref{quanttable}. The plot is not exactly symmetric under $\chi \to -\chi$ due to same random realisation of inhomogeneous modes being used for each run, which breaks the symmetry.}
	\label{deltaN}
\end{figure}

Figure~\ref{deltaN} compares the $\delta N$ functions obtained for three different values of the non-minimal coupling $\xi$. Particularly relevant are the values $\xi = 0$ which corresponds to the minimal coupling case and $\xi = 0.004$ which is the lower bound from observations Eq.~\eqref{xireq}. The same inhomogeneous modes are used to initialise simulation for all $\chi_\text{ini}$. We expect the uncertainty arising from different random realisations of the inhomogeneous modes to be largely uncorrelated between different Hubble volumes, making its effect subdominant to the statistical error in sampling of $\chi_\text{ini}$. As a quantitative check we carry out 32 runs with different random realisations of the inhomogeneous modes for $g^2/\lambda = 2$ and $\xi=0.004$ at $\chi_{\rm ini} \approx 1.581 \times 10^{-6} M_P,  1.203 \times 10^{-6} M_P$ and $0 M_P$. We find the standard deviation in $\delta N$ due to this effect is approximately $40\%$ of the spike heights in Figure~3. Therefore the spikes are a genuine feature of our simulations and this effect is subdominant.

\begin{table}[t]
\centering
\begin{tabular}{|c|c|c|c|c|} \hline
 $\xi$ & $\lambda$ & $\phi_\text{ini}$ & Initial Variance & Cosmic Variance \\ \hline
 $0$ & $1.769 \times 10^{-13}$  & $3.266 M_P$ & $1.044 \times 10^{-12} M^2_P$ & $\infty$ \\
 $0.004$ & $4.946 \times 10^{-13}$  & $3.314 M_P$ & $4.176 \times 10^{-13}M^2_P$ & $2.205 \times 10^{-15} M^2_P$\\
 $0.007$ & $7.334 \times 10^{-13}$ & $3.351 M_P$  & $3.481 \times 10^{-13}M^2_P$ & $5.757 \times 10^{-17} M^2_P$ \\
 \hline
\end{tabular}
\caption{Numerical values of quantities to be specified in simulations for different $\xi$ but all with the same $g^2/\lambda$ = 2.}
\label{quanttable}
\end{table}

Even though the cosmic variance $\langle\overline{\chi}^2\rangle$ is small, it is not exactly zero, and therefore it is important to consider the full range of possible values of $\overline{\chi}$. Because the relevant values are small, we can achieve this without carrying out new simulations by re-weighting the data: For a general small value of $\overline{\chi}$, the coefficients (\ref{NPcoeff})
are given by
\begin{align}
\tilde{N}_\chi &= \frac{1}{N_{\rm runs}}\sum_i 
\exp\left(-\frac{\overline{\chi}^2-2\overline{\chi}\chi_i}{2\langle\delta\chi^2\rangle}\right)
\left(\chi_i-\overline{\chi}\right)\left(N(\chi_i)-\overline{N}\right),
\nonumber\\
\tilde{N}_{\chi\chi} &= \frac{1}{N_{\rm runs}}\sum_i 
\exp\left(-\frac{\overline{\chi}^2-2\overline{\chi}\chi_i}{2\langle\delta\chi^2\rangle}\right)
\left(\chi_i-\overline{\chi}\right)^2\left(N(\chi_i)-\overline{N}\right).
\end{align}

\begin{figure}[t]
    \centering
    \includegraphics[scale=0.33]{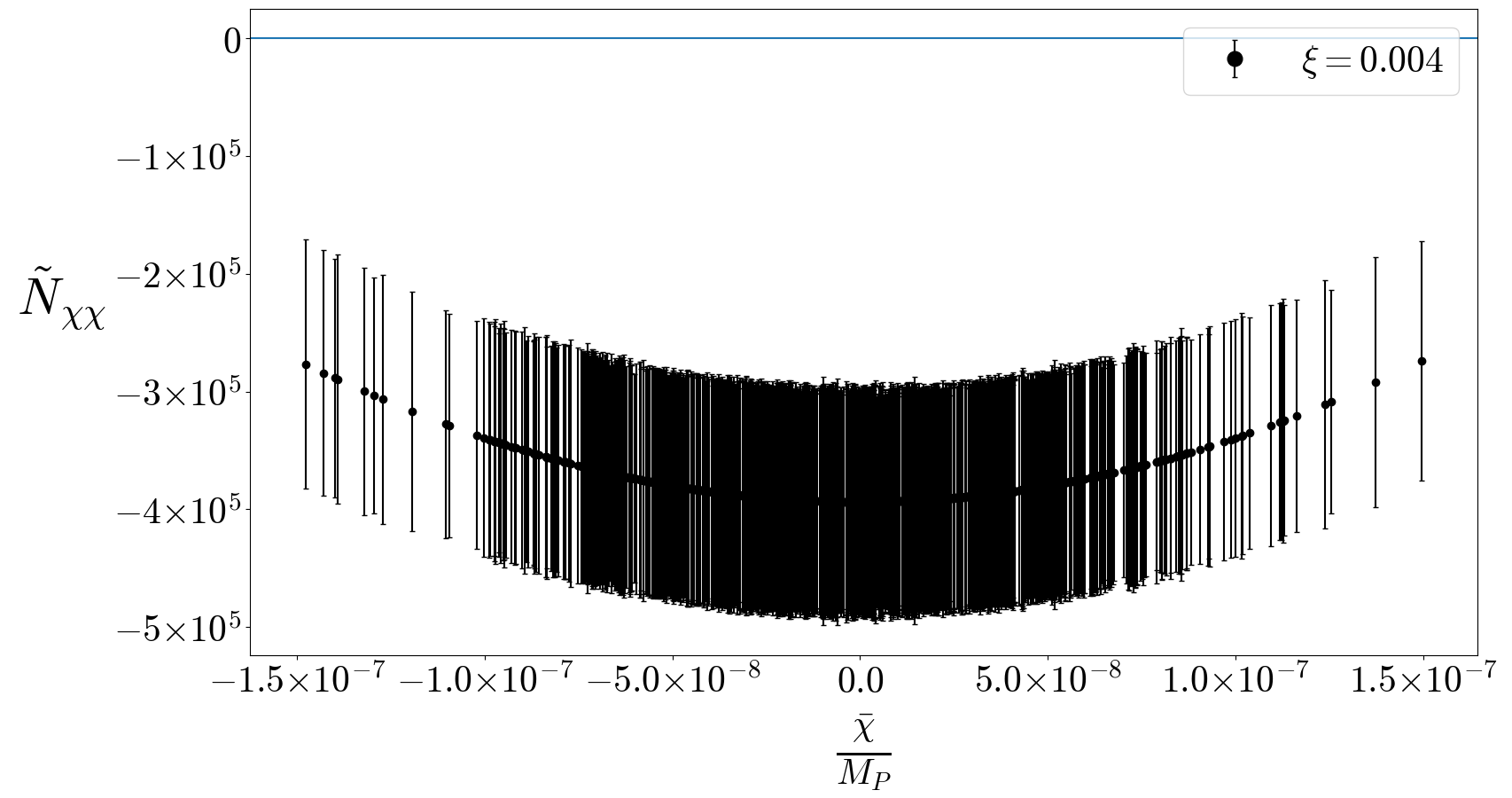}
    \caption{$\tilde{N}_{\chi\chi}$ for 1000 $\overline{\chi}$ drawn from a Gaussian distribution with mean equal to zero and variance equal to cosmic variance $\corr{\overline{\chi}^2}$. The model parameters used are $g^2/\lambda=2$ and $\xi=0.004$. The final value and error bar for each $\overline{\chi}$ are obtained by bootstrapping 100 resamples.}
    \label{G2Ncc}
\end{figure}

Figure~\ref{G2Ncc} shows the non-perturbative coefficient $\tilde{N}_{\chi\chi}$ over the range of $\overline{\chi}$ allowed by the cosmic variance. The final value of $\tilde{N}_{\chi\chi}$ and error bar for each $\overline{\chi}$ are obtained by using the bootstrap (resampling with replacement) method. We use the symmetry $\chi \to -\chi$ to double the number of sample points $\delta N(-\chi)$ used in calculating $\tilde{N}_{\chi\chi}$.

\subsection{Calculating $f_\text{NL}$}
\label{fNLest}

In section \ref{BLapp}, we described how a scale dependent way of solving momentum integrals can significantly change the non-Gaussianity calculated from preheating. Here, we illustrate it using our NMC preheating model.

In a scale invariant treatment, $f_\text{NL}$ is given by Eq.~\eqref{BLfNL}.
Substituting in typical values from our model: $\xi = 0.004$, $\mathcal{P}_\chi \sim 10^{-11}M^2_P$. $\mathcal{P}_* = 2.1 \times 10^{-9}$ is known from Planck observations \cite{Planck:2018nkj}. Taking $L$ to be the length of the observable universe, we have $\ln(k_{*}L) \sim 1$. From lattice simulations, $\tilde{N}_{\chi\chi} \sim 10^6 M^{-2}_P$ for $g^2/\lambda = 2$ and $\xi = 0.004$. Putting these values together gives
\begin{align}
    f^\text{scale inv.}_\text{NL} \sim 1, 
\end{align}
in the scale invariant approximation.

However, the correct $f_\text{NL}$ shall be obtained by taking a scale-dependent estimate of the field variance. For our model of NMC preheating, this essentially implies including the over-damped oscillator contribution $F(\mathcal{N})$ from Eq.~\eqref{NMCPower} in the variance. To obtain $\mathcal{N}$ in terms of $k$, we need to invert the horizon crossing relation $k = a(\mathcal{N}) H(\mathcal{N})$,
\begin{align}
    k = e^{-\mathcal{N}}H(\mathcal{N}) = e^{-\mathcal{N}} \left(\frac{M_P}{\xi} \sqrt{\frac{\lambda}{12}} \left(\frac{16\xi \mathcal{N} +\sqrt{1+32\xi}- 1}{16 \xi \mathcal{N} + \sqrt{1+32\xi} +1}\right)\right).
\end{align}
This is difficult to invert analytically. However, for large $\mathcal{N}$ we notice that the expression is dominated by the exponential factor. Therefore in order to have an analytical estimate, we use the approximation $H = const$ that allows us to capture the large $\mathcal{N}$ behaviour. Fixing $H$ from Eq.~\eqref{Heq} to $H_*$ 
\begin{align}
     \sqrt{\frac{\lambda}{12}} \frac{M_P}{\xi} 
 \frac{16\xi {\cal N}_k +\sqrt{1+32\xi}- 1}{16 \xi {\cal N}_k + \sqrt{1+32\xi} +1} = H_* 
\end{align}
and identifying it in the scale-dependent power spectrum at initial time slice \eqref{NMCPower}
\begin{align}
     \mathcal{P}_\chi(k;t_\text{ini}) \approx \frac{H_*^2}{4\pi^2} e^{-3F({\cal N}_k,{\cal N}_\text{ini})}
\end{align}
with the exponential damping coefficient \eqref{Feq} becoming
 \begin{align}
  F({\cal N}_k,{\cal N}_\text{ini}) &\approx \left(1 - \frac{1}{3}\sqrt{9 - 48\frac{g^2}{\lambda}\sqrt{\frac{\lambda}{12}} \frac{M_P}{H_*}} \right){\cal N}_k  \nonumber\\
 & - {\cal N}_\text{ini} - \frac{\sqrt{32\xi + 1} - 1}{48\xi}\sqrt{9 - 48\frac{g^2}{\lambda}\sqrt{\frac{\lambda}{12}} \frac{M_P}{H_*}}
 \nonumber\\
 &+\frac{2g^2/\lambda}{\sqrt{9 - 48g^2\xi/\lambda}} \tanh^{-1}\left(\sqrt{\frac{3\lambda - 16g^2 \sqrt{\frac{\lambda}{12}} \frac{M_P}{H_*}}{3\lambda - 16g^2\xi}} \right),
 \end{align}
the correlator takes the form
\begin{align}
    \Sigma(q) \approx 2\pi^2 \frac{\mathcal{A}_\chi}{q^{3-n_\chi}},
\end{align}
where
\begin{align}
    n_\chi &= 3 - \sqrt{9 - 48\frac{g^2}{\lambda}\sqrt{\frac{\lambda}{12}}\frac{M_P}{H_*}},\\ 
    \mathcal{A}_\chi &= \frac{H^{2-n_\chi}_*}{2} \exp\left(3\mathcal{N}_\text{ini} +\frac{\sqrt{32\xi+1} - 1}{16\xi}
    \left(3-n_\chi\right)
- \frac{6{g^2}/{\lambda}}{\sqrt{9 - 48\frac{g^2}{\lambda}\xi}} 
    \tanh^{-1}
    \frac{3-n_\chi}{\sqrt{9 - 48\frac{g^2}{\lambda}\xi}}
    \right).
\end{align}
This form separates the momentum $q$ dependence explicitly and is precisely the form \eqref{analVar} we used in Section \ref{BLapp} to calculate an analytical estimate \eqref{scaledfNL} for $f_\text{NL}$.

For typical value of parameters $g^2/\lambda = 2$ and $\xi = 0.004$ in our preheating model, we have $n_\chi \sim 0.1$, $\mathcal{A}_\chi \sim 10^{-12}M^{2-n_\chi}_P$. $\tilde{N}_{\chi\chi} \sim 10^6 M^{-2}_P$ for $g^2/\lambda = 2$ and $\xi = 0.004$ from numerical simulations remains the same. However there is a small subtlety in specifying the value of comoving scale $k_*$. We have fixed scale factor to be one at the end of inflation, $a_\text{end} = 1$. However the physical scale $k_\text{phys} = 0.05 \text{MPc}^{-1}$ at which Planck satellite measurements take place assumes scale factor is one at present times, $a_0 = 1$. Therefore $k_* = (a_0/a_\text{end})k_\text{phys}$. Assuming instantaneous reheating \cite{Dai:2014jja},
\begin{align}
\ln\left(\frac{a_0}{a_\text{end}}\right) = \frac{1}{4} \ln\left(\frac{30}{g_\text{eff} \pi^2}\right)+\frac{1}{3} \ln\left(\frac{11}{43} g_\text{eff}\right) + \ln\left(\frac{\rho_\text{end}^{1/4}}{T_0}\right) \approx 65,
\end{align}
with $g_\text{eff} \sim 100$, $T_0 = 2.725K$ and 
\begin{align}
\rho_\text{end} = \frac{3\lambda}{8} \left(\frac{M_P}{\sqrt{\xi}} \tanh\left(\frac{\sqrt{\xi}}{M_P} \phi_\text{end}\right)\right)^4.
\end{align}
Therefore $k_* \sim 2.225 \times 10^{-30}M_P$. Putting all the above values together, our calculation gives
\begin{align}
    f_\text{NL} \sim 10^{-9} .
\end{align}

The order of magnitude estimate for $f_\text{NL}$ can be made more precise by calculating the momentum integrals involved in Eq.~\eqref{NPfNL} numerically. We split the domain of integration for the momentum $q$ into $q< k_*$ and $q>k_*$. For $q<k_*$, the constant $H_*$ approximation is valid and momentum integral can be obtained analytically as in Eq.~\eqref{scaledfNL}. For $q>k_*$, we invert $q(N)$ numerically to give $\Sigma^\text{num}(q)$ and momentum integral as
 \begin{align}
         \int_{k_*}^{q(\mathcal{N}_\text{ini})} d\vec{q}~\Sigma(q)\Sigma(|\vec{q}-\vec{k}_1|)\Sigma(|\vec{q}+\vec{k}_3|) \approx 4\pi \int_{k_*}^{q(\mathcal{N}_\text{ini})} q^2 ~dq~(\Sigma^\text{num}(q))^3 .
\end{align}
Combining the integrals for the above two domains yields the full integral from $q=0$ to $q=q(\mathcal{N}_\text{ini})$. The remaining ingredient to be used in Eq.~\eqref{NPfNL} is the non-perturbative coefficient $\tilde{N}_{\chi\chi}$. From the lattice simulation results at different $\overline{\chi}$, figure~\ref{G2Ncc}, we take the average $\tilde{N}_{\chi\chi}$ and average error $\Delta \tilde{N}_{\chi\chi}$ to estimate $f_\text{NL}$ and its error through standard propagation. For $g^2/\lambda=2, \xi = 0.004$ we find
\begin{align}
\label{equ:fNL2}
    f_\text{NL} = (2.3 \pm 1.7) \times 10^{-9} ,
\end{align}
which matches well our order of magnitude estimate above.

We can see that there is a significant decrease in the non-Gaussianity estimate if we perform a complete linearised calculation, rather than using the scale invariant approximation (\ref{equ:scaleinv}). Essentially, this occurs because the momentum integral in Eq.~\eqref{NPfNL} is power law suppressed in $k_{*} \sim 10^{-30}M_P$ which results in a very low overall factor in $f_\text{NL}$. Thus we illustrate that for non-Gaussianity calculation in any model of preheating, it is important to include scale-dependence during inflation to yield an accurate result. 

\subsection{Parameter dependence}

\begin{figure}[t]
    \centering
    \includegraphics[scale=0.17]{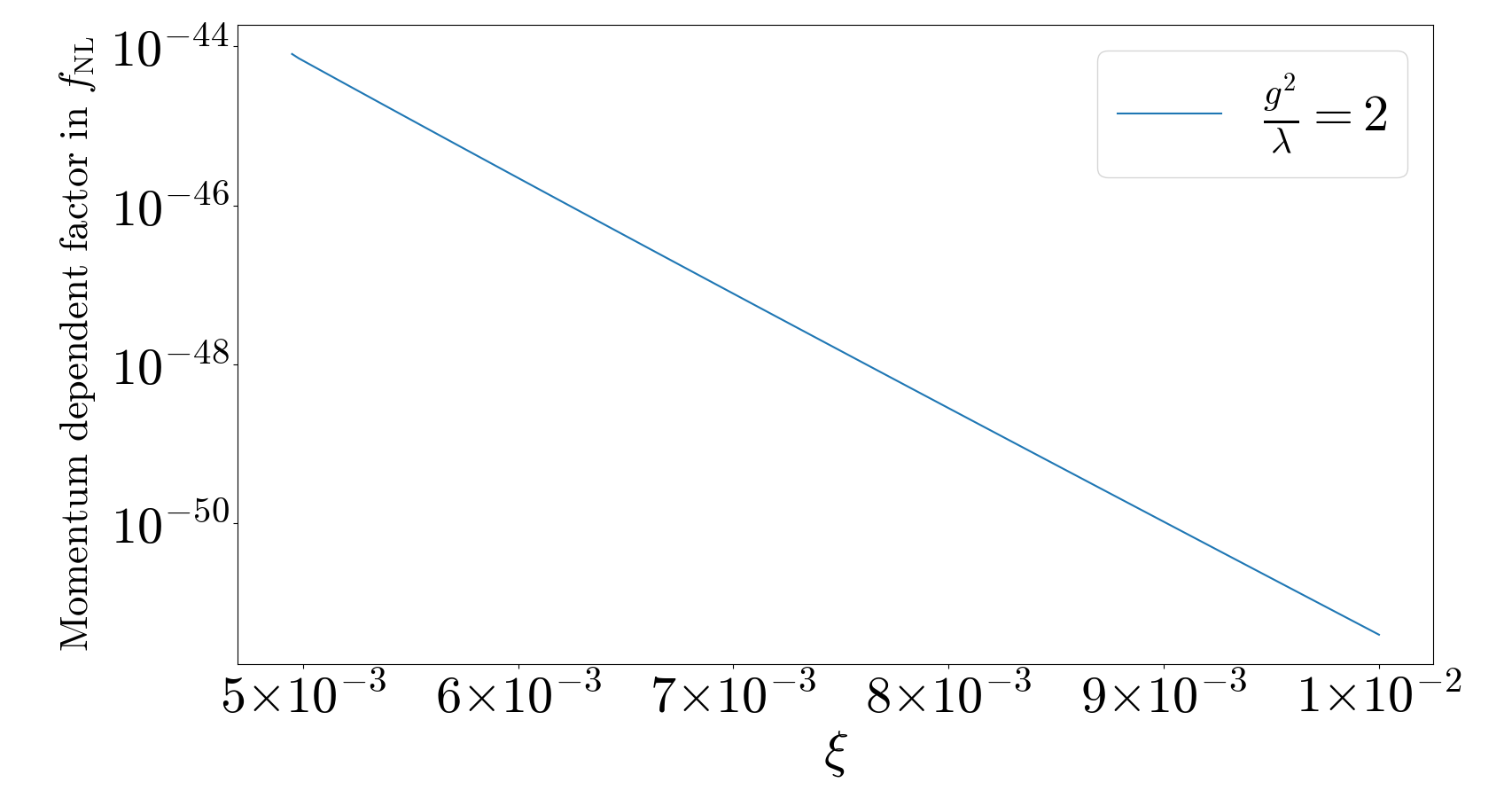} \includegraphics[scale=0.17]{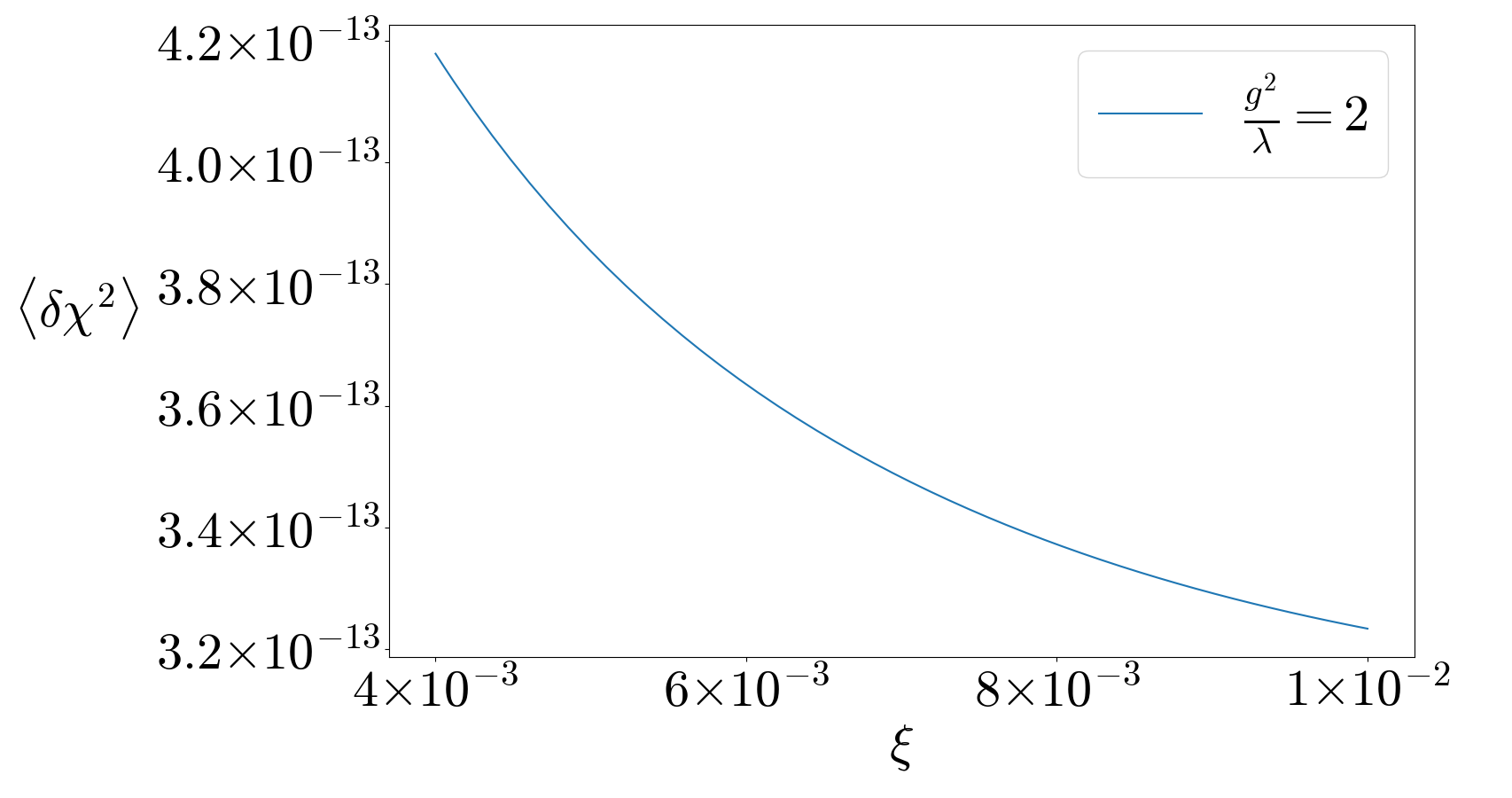}
    \caption{$\xi$ dependence of the momentum integral (left) and initial variance (right).}
    \label{xidep}
\end{figure}

The $\xi$ dependence of momentum integral occurs through the factor $\mathcal{A}^3_\chi k^{3n_\chi}_*/n_\chi$ in Eq.~\eqref{scaledfNL}. Figure~\ref{xidep} shows that this factor drops down almost exponentially as $\xi$ is increased from the lowest value of $\xi=0.004$ allowed in our model. 
The non-perturbative coefficient $\tilde{N}_{\chi\chi}$ Eq.~\eqref{NPcoeff} captures the departure from a constant $\delta N$ within a spread of $\chi_\text{ini}$ governed by the initial variance. Now as can be seen from Figure~\ref{deltaN}, the height of spikes reduces as $\xi$ increases. Furthermore, the number of spikes decreases as the spread in $\chi_\text{ini}$ shrinks. Figure~\ref{xidep} also shows that the initial variance does not change significantly with $\xi$ when compared to the change in momentum integral. Combining both the factors, we surmise $f_\text{NL}$ will be maximum for lowest observational bound on $\xi$, $\xi = 0.004$ for a given value of $g^2/\lambda$.

\begin{figure}[t]
    \centering
    \includegraphics[scale=0.33]{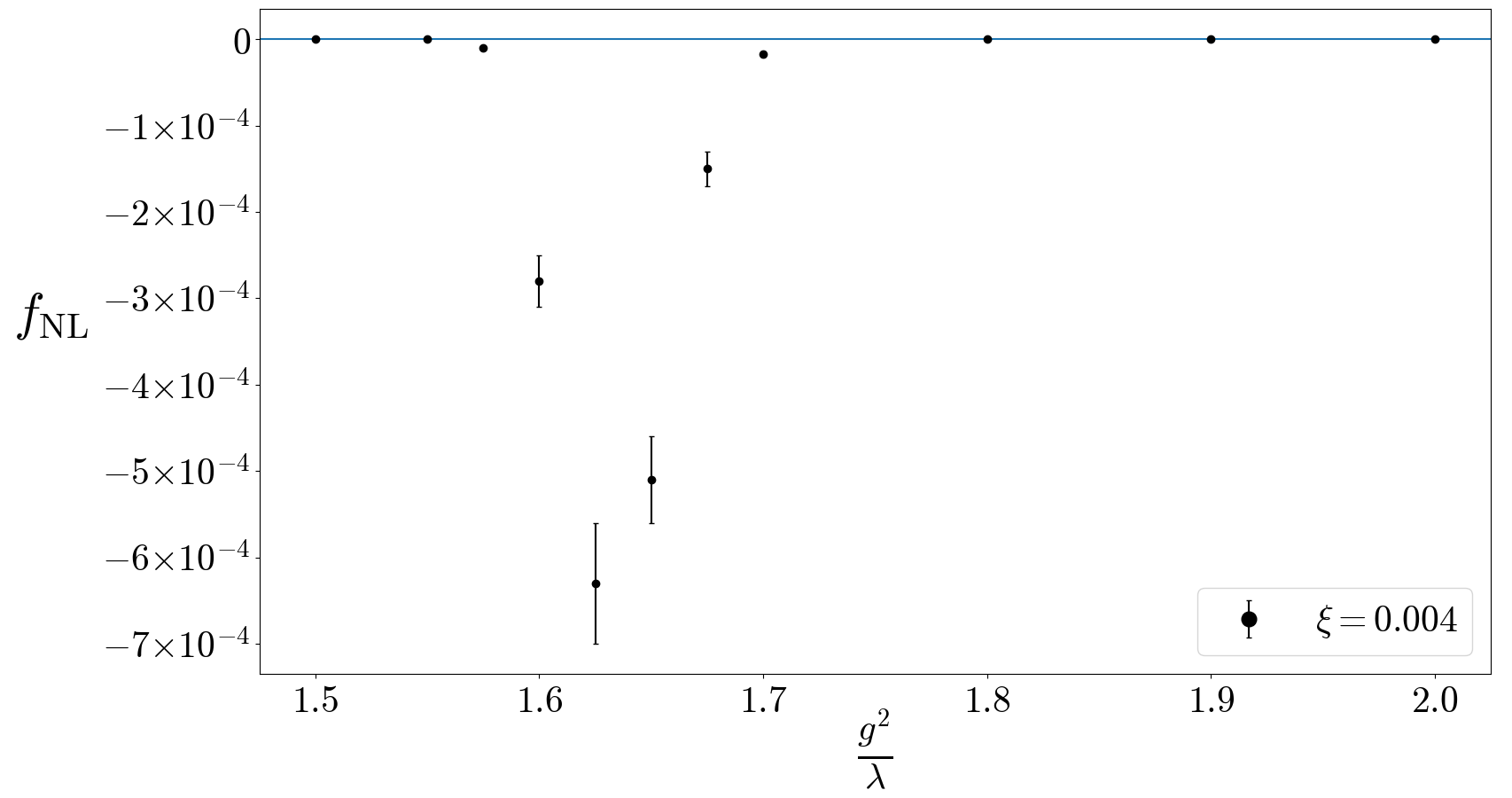}
    \caption{Non-Gaussianity parameter $f_\text{NL}$ at different coupling strengths $g^2/\lambda$ for a fixed $\xi = 0.004$.}
    \label{fNLvsG}
\end{figure}

We can therefore search for the highest value of $f_\text{NL}$ in our model by keeping $\xi = 0.004$ fixed and scanning over different $g^2/\lambda$. Figure~\ref{fNLvsG} shows that the maximum value of $f_\text{NL}$ occurs around $g^2/\lambda = 1.625$. 
At exactly $g^2/\lambda = 1.625$ we find it to be
\begin{align}
f_\text{NL} = -(6.3 \pm 0.7) \times 10^{-4}. 
\end{align}
This value still lies well-within the current observation bound of $f^\text{local}_\text{NL} = -0.9 \pm 5.1$ \cite{Planck:2019kim}, but it is several orders of magnitude larger than what we obtain for generic $g^2/\lambda$, e.g., Eq.~(\ref{equ:fNL2}). 

Based on a linearised analysis of the parametric resonance for the massless preheating case, one might expect the maximum value of $f_\text{NL}$ to lie at $g^2/\lambda = 1.875$ \cite{Chambers:2008gu}. However, our simulations indicate this is shifted to $g^2/\lambda \approx 1.625$. 

\begin{figure}[t]
    \centering
    \includegraphics[scale=0.33]{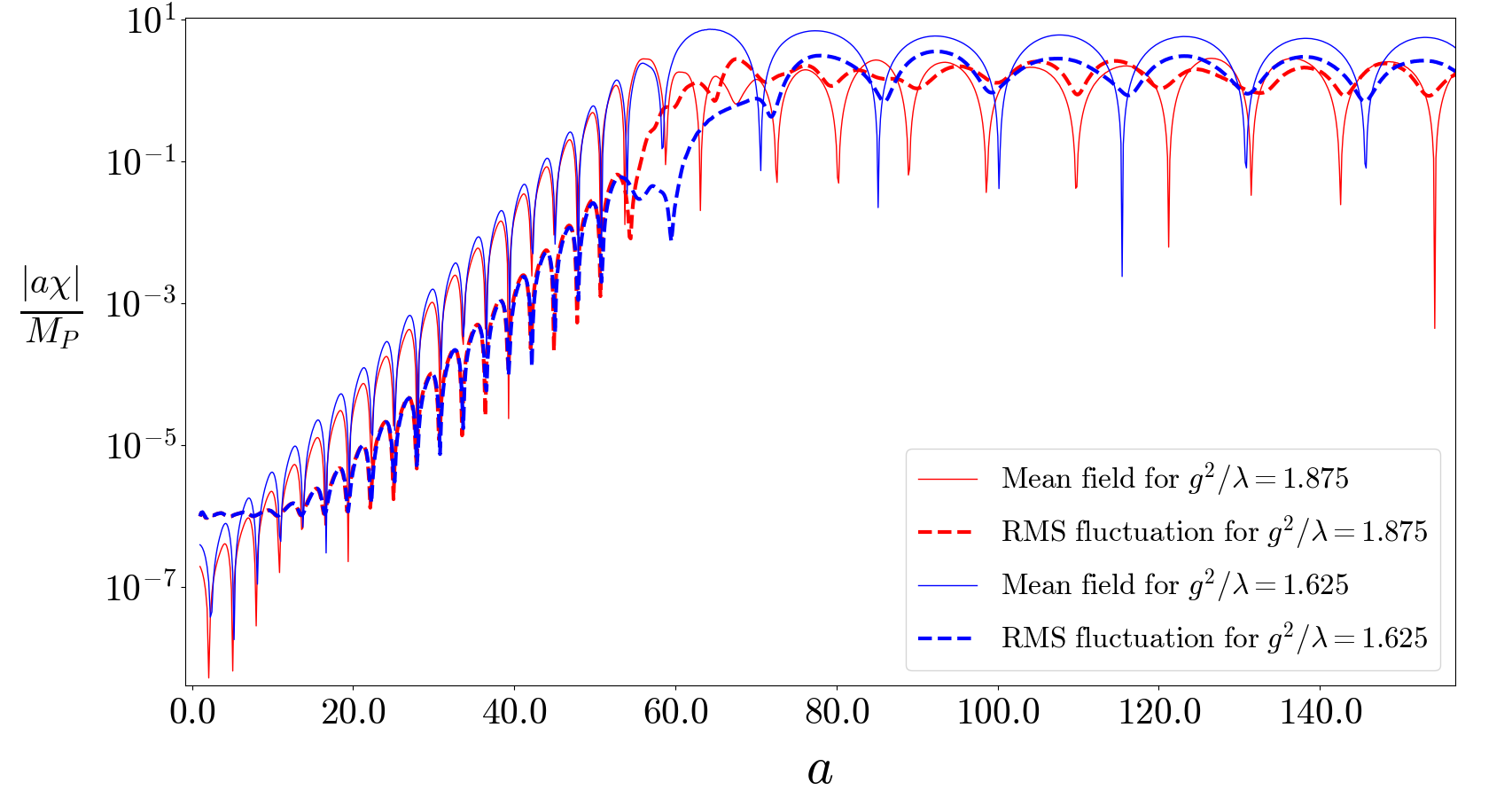}
    \caption{Mean field over all lattice sites $\corr{\chi}$  and root mean squared (RMS) field fluctuation $\sqrt{\langle \chi^2\rangle -\langle\chi\rangle^2}$ values as a function of scale factor for $g^2/\lambda = 1.875$ and $g^2/\lambda = 1.625$. $\xi = 0.004$ in both cases.}
    \label{PRScompare}
\end{figure}

To investigate this shift we can compare the growth of zero mode with root mean square of the inhomogeneous fluctuations in Figure~\ref{PRScompare}. We see that the zero mode for $g^2/\lambda = 1.875$ grows with a slightly higher slope than $g^2/\lambda = 1.625$ in accordance with parametric resonance considerations. However the major difference in evolution for the two coupling strengths arises from the non-linear behaviour of fluctuations. Figure~\ref{PRScompare} shows that after the evolution becomes non-linear, the fluctuations catch up significantly slower for the $g^2/\lambda = 1.625$ value, thus allowing for exponential growth of zero mode to persist for a longer time and hence increasing the $\delta N(\chi)$ that increases $f_\text{NL}$ as compared to $g^2/\lambda = 1.875$ value. This effect can only be captured by lattice simulations. Thus by finding many orders of magnitude change in $f_\text{NL}$ with the coupling strength, we have also demonstrated the need for full lattice simulations to study parameter dependence of preheating models. 

\section{Conclusion and Discussion}

We have extended the non-perturbative formalism \cite{Chambers:2007se,Imrith:2018uyk,Imrith:2019njf} used to calculate non-Gaussianity generated during preheating and illustrated it by applying to an observationally viable model of massless preheating motivated by non-minimal coupling to gravity. 

Interestingly, we find that calculating non-Gaussianity from preheating at currently observed scale, which occurs after the end of inflation, involves taking into account the evolution of modes at very early times, well-before they left the horizon.

In a separate universe picture, the evolution of each individual causal volume of our currently observable universe depends sensitively on initial conditions \cite{Bond:2009xx}. However, the overall probability distribution of initial conditions is Gaussian and depends on the initial variance as well as the mean over our currently observable universe. This mean itself depends on the cosmic variance over the entire universe, of which our observable universe is just a small part. 

If the spectrum of the spectator field is sufficiently blue-tilted during inflation, then the cosmic variance becomes negligible and hence the dependence on modes before the currently observed scale left the horizon disappears. As an example we have shown how this occurs in our preheating model. We are able to arrive at this finding precisely because we have considered the full time-dependence of Hubble rate during inflation. 

When time-dependence of Hubble rate is included, it translates to the spectator field having a scale-dependent spectrum. Thus we may no longer use a simple scale-invariant approximation to calculate non-Gaussianity \cite{Boubekeur:2005fj}. We have also shown how considering scale dependence yields a $f_\text{NL}$ value that is many orders of magnitude different from just using a scale-invariant approximation.

As noted in the text, we drop the effects of non-canonical kinetic terms arising from conversion of Jordan to Einstein frame in the two-field case to simplify the simulations. Simulations with non-canonical kinetic terms have been performed \cite{Nguyen:2019kbm} for non-minimal coupling parameter $\xi > 1$. Such simulations, which incorporate the curvature in field space, are able to capture the full effects of non-minimal coupling and can be used to calculate non-Gaussianity using the formalism outlined in this article. Lattice simulations of preheating with non-minimal coupling can also be performed directly in the Jordan frame \cite{Figueroa:2021iwm}. It would be interesting to perform these simulations for the non-minimally coupled massless preheating model we have considered and compare our results. 

We have scanned for $f_\text{NL}$ over a parameter range $g^2/\lambda = 1$ to $3$ in our model, which is relevant from parametric resonance considerations \cite{Kofman1997}. We find that $f_\text{NL}$ changes by many orders of magnitude within this range. Therefore both detection or non-detection of non-Gaussianity can be used to constrain the coupling strength in preheating models. The maximal $f_\text{NL} \sim 10^{-4}$ occurs around $g^2/\lambda = 1.625$ with $\xi=0.004$ fixed. This value of the non-Gaussianity parameter is consistent with current observational bounds. It is too small to be detected by cosmological experiments in the near future, but perhaps is large enough that a different measure of non-Gaussianity, for example using wavelet transform \cite{Cheng:2020qbx}, can be used to detect it. Finding such a measure that captures the inherently non-linear nature of the fields during preheating is a possible direction for future work.

Nonetheless, the full calculation of $f_\text{NL}$ in our model enables a complete illustration of our formalism to obtain non-Gaussianity from preheating. This method can very well be applied to other interesting models of inflation and reheating, possibly yielding high, detectable non-Gaussianity.    

\acknowledgments

P.S.G. acknowledges financial support from the Government of Maharashtra, India. A.R. was supported by STFC grants ST/T000791/1  and ST/X000575/1 and IPPP Associateship. We wish to thank Zhiqi Huang for the use of the program HLattice (\url{https://www.cita.utoronto.ca/~zqhuang/hlat/}). Lattice simulations were performed using Imperial College London's High Performance Computing facility \cite{HPC}.

\bibliographystyle{JHEP}
\bibliography{JCAPsub.bib}

\end{document}